\documentclass{aa}

\usepackage{graphicx}
\usepackage{subcaption}
\usepackage{txfonts}
\usepackage{hyperref}
\usepackage{comment}
\usepackage{lineno}
\usepackage{multirow}
\usepackage{booktabs}
\usepackage{threeparttable}
\usepackage{xcolor}
\usepackage[normalem]{ulem}

\usepackage{natbib}
\bibpunct{(}{)}{;}{a}{}{,} 

\hypersetup{
    colorlinks=true,                
    breaklinks=true,                
    urlcolor= red,                  
    linkcolor= blue,                
    citecolor= blue                 
    }

\defcitealias{Fichet_2025}{FdC25}

\begin{document} 

\title{Stellar mass loading drives dissipation and reacceleration in AGN jets: Explaining VLBI-\textit{Gaia} offsets and constraining jet power}

   \titlerunning{Stellar mass loading drives dissipation and reacceleration in AGN jets}

   \author{G. Fichet de Clairfontaine
          \inst{1},
          M. Perucho
          \inst{1, 2},
          J. M. Martí
          \inst{1, 2}
          \and 
          Y. Y. Kovalev
          \inst{3}
          }

    \authorrunning{Fichet de Clairfontaine et al.}

   \institute{Departament d’Astronomia i Astrof\'isica, Universitat de Val\`encia, Av/ Vicent Andr\'es Estell\'es, 19, E-46100 Burjassot, Val\`encia\\
              \email{gaetan.fichet@uv.es}
         \and
             Observatori Astron\`omic, Universitat de Val\`encia, C/ Catedr\`atic Jos\'e Beltr\'an 2, E-46980 Paterna, Val\`encia, Spain 
         \and 
            Max-Planck-Institut für Radioastronomie, Auf dem Hügel 69, 53121 Bonn, Germany
             }

   \date{Accepted for publication in A\&A}

\abstract
{Recent Very Long Baseline Interferometry (VLBI) and Gaia astrometry reveal systematic milliarcsecond-scale offsets between the radio and optical centroids of active galactic nuclei (AGN). These “radio-optical offsets” do not alter the standard opacity-driven interpretation of radio core shifts. Instead, they indicate that the optical emission centroid is frequently displaced downstream of the radio synchrotron optical depth $\tau = 1$ surface, implying that additional dissipation and particle reacceleration occur beyond the opacity radio core within relativistic jets.}
{We investigate whether energy dissipation via stellar wind mass-loading (a process by which stellar wind material is entrained into the jet flow, progressively increasing its inertia and triggering dissipation) can generate such offsets via refreshing the particle spectrum, and how this effect may depend on AGN jet power, host stellar distribution, and viewing angle. In addition, another goal of this work is to assess whether radio-optical offsets can serve as a diagnostic of jet composition and energetics across time.}
{We perform steady-state, axisymmetric relativistic magnetohydrodynamic (RMHD) simulations of AGN jets, including baryonic mass-load from stellar winds, varying jet kinetic power, and stellar core radius. Synthetic synchrotron emission maps in radio and optical bands are generated via a radiative transfer code, and centroid offsets are extracted for comparison with observations.}
{Parsec-scale radio-optical offsets arise only for jet powers $L_{\rm j}\sim10^{42.5}-10^{44}\,\rm erg\,s^{-1}$. In this regime, stellar winds trigger jet deceleration at intrinsic distances of a few $10^{2}-10^{3}\,\rm{pc}$, shifting the optical centroid downstream and producing offsets of $\sim0.1-4\,\rm{mas}$ (a few tens of parsecs at $z=1$). Offsets depend on stellar distribution, viewing angle, and optical jet dominance, and vanish outside this power range. We reproduce the observed redshift evolution of offset incidence, linking it to the cosmic evolution of thermally pulsing asymptotic giant branch (TP-AGB) mass loss.}
{Radio-optical offsets provide a flux-weighted, geometry-dependent probe of jet dissipation and stellar entrainment. Although stellar mass loading is unlikely to be the sole dissipation mechanism, its unavoidable presence in galactic nuclei makes it a natural baseline for energy dissipation. Radio-optical offsets therefore offer a constraint on AGN jet power and jet-host coupling, independent of traditional lobe-based methods.}

   \maketitle
%

\section{Introduction}
\label{sec:Introduction}

A significant fraction of active galactic nuclei (AGN) launch collimated, relativistic jets that can extend from parsec to megaparsec scales, transporting enormous amounts of energy, up to $\sim10^{46}~\rm{erg\,s^{-1}}$ in kinetic power \citep{Blandford_2019}, from the vicinity of the supermassive black hole into the host galaxy and beyond. These jets play a central role in galaxy evolution and feedback, regulating star formation and heating cluster gas \citep{mcnamara_2007,Cavagnolo_2010, Fabian_2012, Blandford_2019}. Among them, radio galaxies host jets misaligned with respect to the observer's line of sight (i.e.\ seen at large angles to the jet axis), which are classified according to the Fanaroff-Riley (FR) scheme \citep{Fanaroff_1974}. This classification divides sources into FRI and FR\,II classes, based on radio morphology and luminosity. FR\,I jets are less luminous and show centrally peaked radio emission, whereas FR\,II jets are brighter and terminate in compact hotspots at the outer edges of their radio lobes
\citep[e.g.,][]{Fanaroff_1974,Bridle_1984}. This was later interpreted in terms of FR\,I jet deceleration via mixing \citep{Bicknell_1984,Bicknell_2002}.
Unified schemes further link these morphological classes to orientation and accretion state, with FR\,I and FR\,II radio galaxies as the misaligned counterparts of BL Lac type objects and radio-loud quasars, respectively.
Despite this phenomenological framework, the physical process responsible for the FR\,I/II dichotomy remains uncertain \citep[e.g.,][]{Bicknell_2002, Wold_2007,Tchekhovskoy_2016,Gourgoliatos_2018,perucho_2019,Perucho_2020}, with key questions about jet energetics and composition still unanswered.

One critical factor in distinguishing FR\,I and FR\,II sources is jet kinetic power. FR\,II jets typically carry higher power ($\gtrsim10^{45}~\rm{erg}\,\rm{s}^{-1}$) than FR\,I jets \citep[e.g.,][]{Blandford_1979,Godfrey_2013}. However, this is not the only intervening parameter; stellar composition and host galaxy properties also play a role, mainly at intermediate powers, as pointed out by, e.g., \citet{Mingo_2022}. In practice, AGN jet power spans many orders of magnitude and is only indirectly inferred, with large uncertainty implied. Power and composition together govern how a jet interacts with its environment: high-power jets can remain collimated and relativistic to large distances, whereas lower-power jets are prone to entrainment and deceleration. However, stability analysis of relativistic flows shows that thermodynamically cold (specific enthalpy dominated by rest-mass energy) and fast (kinetically relativistic) jets are intrinsically more stable \citep{Perucho_2005, Perucho_2010} and whereas powerful jets tend to have more inertia (or mass flux), these properties do not necessarily need to be correlated, mainly at compact (sub-kiloparsec) scales. Still, all these quantities remain poorly constrained by observations.

As mentioned above, estimates of the AGN jet kinetic power remain challenging due to large systematic uncertainties \citep[e.g.,][]{Ineson_2017, Hardcastle_2020}. Common methods either use scaling relations between radio synchrotron luminosity and kinetic power \citep{Godfrey_2016, Croston_2018}, or derive the work done in inflating X-ray cavities \citep{McNamara_2012, Pasini_2021}. Radio-based estimates depend on assumptions about particle content, magnetic field strength, spectral shape, and source age, and may vary by orders of magnitude depending on the presumed energy partition or proton content \citep{Godfrey_2016}. X-ray cavity measurements offer a more direct probe of the jet energy budget but require well-resolved structures and assume that the observed cavities fully trace jet energy with minimal leakage \citep{McNamara_2012}, plus an estimate of the jet age that may itself be uncertain by more than one order of magnitude. Both methods provide time-averaged powers and may miss effects such as episodic activity, which is increasingly observed in remnant and restarted radio sources as the sensitivity of arrays increases \citep{Brienza_2017, Shabala_2020}. 

The composition of jets is equally elusive but critically important for jet dynamics. Jets might be made of electron-positron pairs or pair-proton plasmas, and the presence of heavy particles greatly increases the inertia of the flow \citep{2002MNRAS.331..615S, Blandford_2019} if they represent a significant fraction of the bulk flow mass flux. Observational probes of composition (such as Faraday rotation or lobe pressures) suggest a range of possibilities \citep[e.g.,][]{Croston_2018} at the large scales but are often compatible with both leptonic and hadronic solutions. The question of jet composition is also intertwined with the potential emission of very energetic neutrinos \citep{Mannheim1993, Murase2014, Murase2023}, as relativistic protons are necessary to produce them. Stars in the jet could be the missing link between mass-loading and PeV neutrino production \citep{Fichet_2025b}. 

Interestingly, FR\,I radio galaxies require a larger thermal population in their lobes than FR\,IIs, which is in agreement with the former being significantly mass-loaded \citep{Croston_2018}. Theoretical studies show that light, pair-dominated jets decelerate strongly as soon as they entrain even a relatively small baryonic load, whereas jets that are already proton-rich carry far more kinetic power per unit radio luminosity and are much less affected by additional mass‐loading \citep[e.g.,][]{Komissarov_1994, Bicknell_1995}. Importantly, jets interact with stars and interstellar medium (ISM) gas as they propagate: stellar winds, interstellar clouds, and small-scale instabilities at the jet surface can load jets with baryonic material \citep[e.g.,][]{Komissarov_1994,Bowman_1996,Perucho_2014,Matsumoto_2017,Gourgoliatos_2018,perucho_2019,Perucho_2020,Angles_2021}. These processes were originally proposed to explain the FR\,I/II divide due to low-power jets being transonic and turbulent through mass-load, whereas high-power jets remain supersonic \citep{Bicknell_1995}. In this picture, mass-load and composition are intimately linked to the large-scale morphology of radio galaxies.

Furthermore, mass loading by stars has also been suggested to be connected to intriguing observational features recently identified. High-precision astrometry reveals small but systematic offsets between the apparent radio core (from VLBI) and the optical centroid (from \textit{Gaia} measurements) in many AGN \citep{Plavin_2019,Petrov_2019}. In particular, \cite{Plavin_2019} found that in most cases, the optical position is displaced downstream along the jet direction by milliarcseconds (mas). This “radio-optical offset” implies that the optical emission centroid is often dominated by extended jet light rather than by the nucleus, in contrast to the expected transition controlled by opacity in the jet. Extended parsec-scale optical jets shifting the Gaia photocenter downstream, contamination by host galaxy optical emission, or accretion disk dominance producing upstream offsets have been proposed as possible explanations \citep{Kovalev_2017, Petrov_2019, Kovalev_2020}.

However, \cite{Fichet_2025} (hereafter \citetalias{Fichet_2025}) showed that stellar mass loading can naturally generate positive radio-optical offsets through the dissipation it induces in the jet if a fraction of the dissipated energy is invested in particle acceleration. In their numerical experiments, jets with FR\,I-like powers ($L_{\rm j}\sim 10^{43}~\rm{erg}\,\rm{s}^{-1}$) entrained mass from realistic stellar distributions, which forced a gradual conversion of kinetic energy into internal energy along the flow. Radiative-transfer calculations with the code \texttt{RIPTIDE} \citep{Fichet_2021} showed that dissipation can increase the minimum Lorentz factor of the non-thermal electron population downstream, causing the optical synchrotron emission to peak farther from the jet base than the radio core. As a result, offsets of a few parsecs arise naturally. Under this assumption, namely that the dissipation-driven increase of $\gamma^\prime_{\min}$ shifts the optical centroid, \citetalias{Fichet_2025} found that even moderate stellar mass-loading rates ($\sim10^{-9}~M_\odot~\rm{yr}^{-1}\,\rm{pc}^{-3}$) 
reproduce the typical offset magnitudes and alignments reported observationally \citep[see also][]{Kovalev_2017}. In essence, the offset is a direct imprint of where dissipation becomes most efficient: the optical emission traces the downstream dissipation zone, while the radio peak remains anchored to the optically thick jet base.

The recent mass-load models connect two previously separate ideas. Historically, stellar entrainment was proposed as a possible explanation for the slow and diffuse nature of FR\,I jets \citep[e.g.,][]{Komissarov_1994, Bowman_1996,Perucho_2014}. We invoke the same physical mechanism to explain radio-optical position shifts. The region in which these interactions could be more intense would coincide with the strong dissipation region identified by \cite{2014MNRAS.437.3405L}. \citetalias{Fichet_2025} suggest that these positional offsets could act as a novel probe of jet physics: they report that jets with the most pronounced positive offsets correspond to moderately low-power FR\,I jets, whereas most powerful jets would tend to show no offset or even negative offset (optical upstream of radio) because they remain relativistic and do not show indications of strong, continuous dissipation at scales of tens to hundreds of parsecs. In other words, the sign and magnitude of the offset depend on the balance of jet power and entrainment. If this interpretation holds, then measuring radio-optical offsets could place independent constraints on jet energetics, complementing traditional methods. 

Motivated by these considerations, we undertake a systematic study of jet dynamics, composition, and resulting emission with a focus on radio-optical offsets. In particular, we extend the work of \citetalias{Fichet_2025}, limited to a single jet power, by exploring a range of jet kinetic powers and host stellar distributions. We parameterize the stellar density profile (using a core radius $r_{\rm c,s}$) to represent different galactic environments. By varying the jet power and mass-loading profile, we aim to quantify the interplay between these factors and their impact on jet deceleration and the relative positioning of radio and optical emission. Our goal is to assess whether observed radio-optical offsets can help constrain the kinetic power and composition of AGN jets, thereby shedding light on the long-standing questions of the FR I/II classification and jet feedback. 

The paper is organized as follows. The Sect.~\ref{sec:Numerical setup} describes our numerical tools and setup. Then, in Sect.~\ref{sec:dynamics} and Sect.~\ref{sec:radiation}, we present our results related to jet dynamics and radiative output, respectively. In Sect.~\ref{sec:discussion}, we discuss these results in the frame of the radio-optical offset observations and other observational evidence before presenting our conclusions in Sect.~\ref{sec:Conclusions}. Throughout this paper, quantities given in the jet co-moving frame are primed, and we assume a flat $\Lambda$CDM cosmology with ${H}_{\rm 0} = 69.6~\rm{km}\,\rm{s}^{-1}\,\rm{Mpc}^{-1}$, $\Omega_{\rm 0} = 0.29$ and $\Omega_{\Lambda} = 0.71$. 

\section{Numerical setup}
\label{sec:Numerical setup}

\subsection{Jet simulation method}
\label{subsec:Jet simulation}

The dynamical simulations are carried out with the quasi-one-dimensional RMHD code described \cite{Angles_2021} and further used in \citetalias{Fichet_2025}. Under the approximations of a narrow jet (jet radius much smaller than its length) and a relativistic flow, these models can be built by solving the time-dependent (magneto-)hydrodynamical equations for the transversal flow with the time coordinate acting as the axial coordinate in the steady flow \citep[quasi-one-dimensional approximation;][]{Komissarov_2015}, with the appropriate boundary conditions at the jet–ambient medium interface. The code uses a Synge equation of state to describe a mixture of ideal, relativistic gases of protons, electrons, and positrons. Jets are injected with leptonic composition (leptonic fraction $\eta\equiv\rho_{\rm e}/\rho=1$), in pressure equilibrium with the ambient medium at the grid boundary \citep[see][]{Marti_2015, Marti_2016}, with constant initial flow density $\rho_{\rm j}$, Lorentz factor $\gamma_{\rm j}$, and axial ($B^{z}_{\rm j}$). The toroidal magnetic field is also fixed, and its profile is shown in \cite{Angles_2021}. The total jet power is defined as
\begin{equation}
    L_{\rm j} = \pi R^2_{\rm j} \left(\rho_{\rm j} h_{\rm j} \gamma^2_{\rm j} + \left(B^\phi_{\rm j}\right)^2 \right) v_{\rm j} \,,
\end{equation}
where $R_{\rm j}$ is the jet radius, $h_{\rm j}$ is the specific enthalpy, and $v_{\rm j}$ is the flow velocity. For reference, we adopt the injection parameters of model J4\_C from \cite{Angles_2021}: a cold leptonic jet with rest-mass density $\rho_{\rm j,0}$, Lorentz factor $\gamma_{\rm j,0}=6$, and injection radius $R_{\rm j}=1~\rm{pc}$. In contrast to \citetalias{Fichet_2025}, where jet power was fixed at $L_{\rm j}=10^{43}~\rm{erg}\, \rm{s}^{-1}$ for an FR\,I-type value, we explore a wide range of jet powers from $10^{41}$ to $10^{46}~\rm{erg}\, \rm{s}^{-1}$ (including both FR\,I and FR\,II regimes). 

Throughout this work, we vary the jet kinetic power solely by modifying the rest-mass density at injection. All other jet parameters (particularly the Lorentz factor, specific enthalpy, and magnetic field strength) are kept fixed. This choice is motivated by the expectation that, at the scales we study, these parameters remain close to the values adopted here, and reaching jet powers of $10^{46}~\rm{erg}\,\rm{s}^{-1}$ by increasing them would result in unrealistic values. Although mixed solutions (involving, e.g., simultaneous increases in rest-mass density and Lorentz factor) are also possible, we adopt the simplest approach to isolate the effect of the rest-mass density on the jet power, and therefore on the radio–optical offset. This choice is further supported by the results of \citet{Angles_2021}, who showed that varying the injection Lorentz factor within the range $\gamma_{\rm j} = 5-10$ \citep[consistent with the typical values observed in parsec-scale jets; e.g.,][]{Lister_2021} does not significantly alter either the deceleration distance or the global impact on the dynamics (see their Fig.~5). Regarding the magnetic field, \citet{Zamaninasab_2014} showed that the jet magnetic flux correlates with accretion power over seven orders of magnitude across the AGN population, implying $B(1~\textrm{pc}) \propto L_{\rm j}^{1/2}$ for a fixed jet cross-section; field strengths inferred from parsec-scale VLBI core-shift measurements \citep{Lobanov_1998} span roughly one to two orders of magnitude across the population. However, by the parsec scales at which our simulations begin, jets have already converted the bulk of their Poynting flux into kinetic energy through magnetic acceleration \citep{Vlahakis_2004, Komissarov_2007, Chatterjee_2019}, transitioning from a magnetically dominated flow near the black hole to a kinetically dominated one. At the tens-of-parsec to kiloparsec scales where stellar mass loading and dissipation operate, the magnetic contribution to the energy flux is most likely negligible compared to the bulk kinetic flux. Hence, fixing $B_{\rm j}$ at injection is thus a simplification, but it is not expected to alter our qualitative conclusions.

The computational grid covers $20~\rm{pc} \times 2000~\rm{pc}$ in cylindrical coordinates $(r,z)$ with $800 \times 5000$ cells, spanning along several stellar core radii \citep[see][]{Angles_2021}. This resolution is adequate to reach convergence, as shown in Appendix\,\ref{app:convergence}. In total, we made $400$~simulations ($100$ simulations with different jet powers $L_{\rm j}$ for two stellar core radii, $r_{\rm c, s} = 0.5, \, 1~\rm{kpc}$, and $100$ simulations with different stellar core radii $r_{\rm c,s}$ for two jet powers $L_{\rm j} = 10^{43}, \, 10^{44} ~\rm{erg}\,\rm{s}^{-1}$). 

We emphasize that the quasi-one-dimensional RMHD framework adopted here is designed to accurately capture the dynamics of relativistic, supersonic flows. In practice, the numerical description remains reliable as long as the jet bulk axial velocity satisfies $v \gtrsim 0.9c$ (corresponding to Lorentz factors $\gamma_{\rm j} \gtrsim 2.3$). Below this threshold, the assumptions underlying the steady, axisymmetric treatment become progressively less accurate, as the flow is expected to enter a strongly decelerated and potentially turbulent regime that cannot be faithfully captured by our present approach. 

\subsection{Ambient medium and mass load}
\label{subsec: Ambient medium and mass loading}

The galactic ISM is modeled by a King-like profile, as in \cite{Angles_2021},
\begin{equation}
    p_{\rm a}\left(z\right) = p_{\rm a, 0} \left[1 + \left(\dfrac{z}{r_{\rm c}}\right)^{2}\right]^{\alpha} \,,
\end{equation}
with $\alpha = - 1.095$, the core radius $r_{\rm c}$, base pressure $p_{\rm a,0}$, and density $\rho_{\rm a,0}$ chosen to ensure pressure balance at the jet base, following \citet{Angles_2021}. In practice, we set $r_{\rm c}=1.2~\rm{kpc}$, as in previous works \citep{Perucho_2014}. Mass loading from stellar winds is introduced via a source-term profile that traces the stellar density. We take,
\begin{equation}
    Q\left(z\right) = Q_{\rm 0} \left[1 + \left(\dfrac{z}{r_{\rm c, s}}\right)^{2}\right]^{\alpha_{\rm s}} \,,
\end{equation}
with $\alpha_{\rm s} = -1.095$, and where $r_{\rm c,s}$ is the core radius of the host’s stellar distribution. Here, $Q(z)$ \citep[in units of ${\rm g\, yr^{-1}\, pc^{-3}}$][]{Bowman_1996} gives the mass injection rate per unit volume at height $z$. In our simulations, we vary $r_{\rm c,s}$ over the range $0.1 - 1.5~\rm{kpc}$ to assess its effect on jet dynamics and emission. This represents another extension of the study of \citetalias{Fichet_2025}, where it was defined in the range $r_{\rm c,s}\in \left[1.0,1.5\right]~\rm{kpc}$. The base loading rate $Q_{\rm 0}$ is chosen to reflect a typical old stellar population in an elliptical galaxy, $Q_{\rm 0}\sim 5 \times 10^{24}~\rm{g}\,\rm{yr}^{-1}\, \rm{pc}^{-3}$ (which corresponds to an average mass-loss rate of $10^{-9} - 10^{-10}~M_\odot \, \rm{yr}^{-1}$ for a stellar density $n_\star \sim 1~\rm{pc}^{-3}$), within the range studied in \citetalias{Fichet_2025}. The stellar winds are assumed to be initially cold (proton-electron plasma) and to be entrained into the jet flow 
\citep[see, e.g.,][]{Bosch_2012,Perucho_2017}. Table\,\ref{tab:param} summarizes the key physical parameters for the jet, ambient medium, and mass loading in our simulation suite.

%
\begin{table*}
\centering
\caption{Simulation parameters. Fixed (fiducial) values are in bold; ranges indicate the grid explored in this study.}
\label{tab:param}
\begin{tabular}{llcccp{4cm}}
\hline\hline 
Category & Parameter & Symbol & Value / Range & Unit & Notes / Reference \\[2pt] \hline \\[2pt]
\multirow{4}{*}{Jet injection} 
 & Nozzle radius & $R_{\rm j}$ & $\mathbf{1}$ & $\rm{pc}$ & As in model J4\_C of \cite{Angles_2021} \\
 & Lorentz factor & $\gamma_{\rm j,0}$ & \textbf{6} & $-$ & kinetically dominated, leptonic jet \\
 & Kinetic power & $L_{\rm j}$ & $10^{41}$-$10^{46}$ & $\rm{erg}\, \rm{s}^{-1}$ & This work (spans FR I-II) \\[2pt]
\hline
\\[2pt]
\multirow{4}{*}{Ambient medium} 
 & Core radius & $r_{\rm c}$ & $\mathbf{1.2}$ & kpc & King profile, as in \cite{Perucho_2014}. \\
 & Pressure at $z=0$ & $p_{\rm a,0}$ & $\mathbf{10^{-7}}$ &$\,\mathrm{dyn\, cm^{-2}}$ & \cite{Angles_2021} \\
 & Density at $z=0$ & $\rho_{\rm a,0}$ & $\mathbf{10^{-24}}$ & $\rm{g}\, \rm{cm}^{-3}$ & $-$ \\
 & Slope & $\alpha$ & $\mathbf{-1.095}$ & $-$ & $-$\\[2pt]
\hline
\\[2pt]
\multirow{3}{*}{Stellar mass loading} 
 & Core radius & $r_{\rm c,s}$ & $0.1$-$1.5$ & kpc & Elliptical host variants \citep{Lauer_2007} \\
 & Slope & $\alpha_{\rm s}$ & $\mathbf{-1.095}$ & $-$ & Mirrors gas profile \\
 & Base rate & $Q_{0}$ & $\mathbf{5 \times 10^{24}}$ & $\rm{g}\, \rm{yr}^{-1}\, \rm{pc}^{-3}$ & Consistent with evolved stellar populations \citep{Vieyro_2017} \\[4pt]
 \hline
\end{tabular}
\end{table*}

\subsection{Radiative transfer with the \texttt{RIPTIDE} code}
\label{subsect: riptide}

We compute synthetic synchrotron emission maps by post-processing our relativistic jet simulations with the \texttt{RIPTIDE} code \citep{Fichet_2021,Fichet_2022}. In this model, the non-thermal electron population in the jet’s co-moving frame is prescribed as a power-law,
\begin{equation}
    \dfrac{\textrm{d}n^\prime_{\rm e}}{\textrm{d}\gamma^\prime_{\rm e}} = K \gamma^{\prime - p}_{\rm e}~~~\textrm{for}~\gamma^\prime_{\rm e,min} < \gamma^\prime_{\rm e} < \gamma^\prime_{\rm e, max} \,,
\end{equation}
with spectral index $p\simeq2.2$ 
\citep[a typical value for mildly relativistic shock acceleration; see e.g. ][]{Ostrowski_2002,Lemoine_2003}. The normalization constant $K$ and the low-energy cutoff $\gamma^\prime_{\rm e,\min}$ are determined from the local non-thermal electron number and energy densities ($n^\prime_{\rm e},e^\prime_{\rm e}$) via the relations of \cite{Gomez_1995},

\begin{equation}
    K = \left(\dfrac{e^\prime_{\rm e} (p-2)}{m_{\rm e} c^2 (1 - C_{\rm E}^{2-p})}\right)^{p-1} \left( \dfrac{1 - C_{\rm E}^{1-p}}{n^\prime_{\rm e} (p-1)}\right)^{p-2}\,,
\end{equation}

\begin{equation}
    \gamma^\prime_{\rm e, min} = \dfrac{1}{m_{\rm e} c^2} \dfrac{e^\prime_{\rm e}}{n^\prime_{\rm e}} \dfrac{p - 2}{p - 1} \dfrac{1 - C_{\rm E}^{1-p}}{1 - C_{\rm E}^{2-p}}\,,
\end{equation}
where $n_{\rm e}^\prime$ is the number density of the nonthermal electrons and $e_{\rm e}^\prime$ its corresponding energy density, and $C_{\rm E}$ is the fixed ratio $C_{\rm E}=\gamma^\prime_{e,\max}/\gamma^\prime_{\rm e,\min} = 10^3$. We consider radiative cooling to be dynamically negligible at the  simulated scales of hundreds of parsecs to kiloparsecs (see \citetalias{Fichet_2025}).

We assume that a fixed fraction of the local thermal energy and leptonic number is transferred to the non-thermal electrons: $e^\prime_{\rm e}=\epsilon_{\rm e} e^\prime_{\rm th}$ and $n^\prime_{\rm e}=\xi_{\rm e} n^\prime_{\rm e, th}$ with $\epsilon_{\rm e}=\xi_{\rm e}=0.1$ \citep[e.g.,][]{Gomez_1995,Mimica_2012,Fromm_2016}. Here $e^\prime_{\rm th}$ and $n^\prime_{\rm e, th}$ are the co-moving internal energy and lepton density from the simulation. Hence, $\gamma^\prime_{\rm e,\min}\propto e^\prime_{\rm th}/n^\prime_{\rm e,th} = \bar{\epsilon}^\prime_{\rm e}$, i.e.\ the mean internal energy per particle. As a consequence, $\gamma^\prime_{\rm e,\min}$ increases in regions of strong dissipation (e.g., shocks or turbulent mixing regions) in which the internal energy is enhanced and efficient particle acceleration can occur. 

This dependence is central to reproducing the observed radio-optical positional offsets (\citetalias{Fichet_2025}). In our model, higher-energy electrons are preferentially produced in regions where jet-star interactions enhance local dissipation, shifting the optical emission centroid downstream relative to the opacity-defined radio core. In contrast, adopting a fixed $\gamma^\prime_{\rm e,\min}$ prescription (e.g., $\gamma^\prime_{\rm e,\min}=1$) suppresses this spatial differentiation: the optical emissivity either remains co-spatial with the radio emission or falls below the observational detection threshold, depending on the local physical conditions. We note that even in the case of intrinsically co-spatial emission, systematic offsets may arise observationally because optical positions correspond to flux-weighted centroids, whereas VLBI radio positions trace the brightest compact core component. In our analysis, centroids are extracted from PSF-convolved synthetic maps without full instrumental forward modeling and therefore represent idealized observational proxies. 

The synchrotron emissivity $j^\prime_{\nu^\prime}$ and absorption coefficients $\alpha^\prime_{\nu^\prime}$ are computed in the co-moving frame at each cell using standard formulae \citep[e.g.,][]{Rybicki_1979, Katarzynski_2001}. These coefficients are then transformed to the observer frame via the Doppler factor $\delta=[\gamma_{\rm j}(1-\beta_{\rm j}\cos\theta_{\rm obs})]^{-1}$ (with the bulk Lorentz factor $\gamma_{\rm j}$ at the cell and line-of-sight angle $\theta_{\rm obs}$), according to,
\begin{equation}
    j_\nu = \delta^2~j^\prime_{\nu^\prime}~~\rm{and}~~\alpha_\nu = \delta^{-1} \alpha^\prime_{\nu^\prime}\,,~~~\textrm{with}~~~\nu = \dfrac{\delta}{1 + z} \nu^\prime \,,
\end{equation}
so that the emission is boosted in the direction of motion. 
We solve the radiative transfer equation $\textrm{d}I_\nu/\textrm{d}\tau_\nu = S_\nu-I_\nu$ (assuming a constant synchrotron source function, $S_\nu = j_\nu/\alpha_\nu$, within the cell) along each line of sight to accumulate the emission and absorption through all the cells \citep{Rybicki_1979}.
Relativistic beaming makes the approaching jet appear significantly brighter and more compact to the observer. We account for the counter-jet by mirroring the emission and deboosting it. However, for the range of observational angle study, the effect of these counter-jets on the radio and optical centroids is expected to be limited \citep[the intensity ratio between the jet and counter-jet is expected to be $\sim 10^3$ and $\sim 10^5$ for $\theta_{\rm obs} = 30^\circ$ and $5^\circ$, respectively; see][] {Ghisellini_1993}. 

\begin{figure*}
    \centering
    \includegraphics[width=\linewidth]{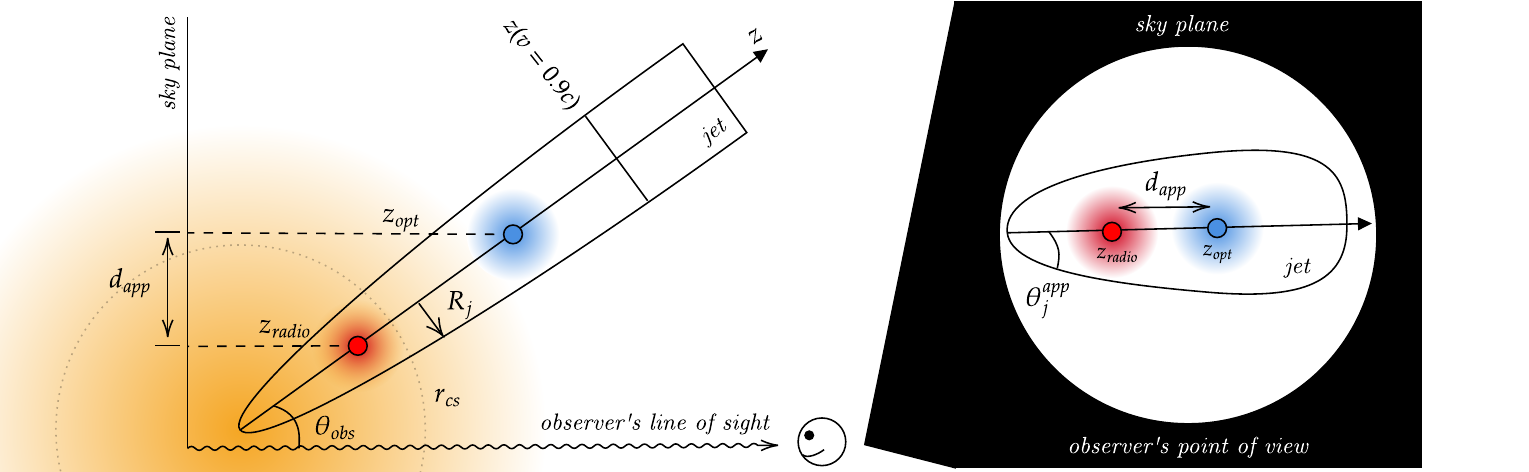}
    \caption{Schematic of the jet geometry. The approaching jet propagates from the nucleus at angle $\theta_{\rm obs}$ to the line of sight. The radio and optical centroids ($z_{\rm radio}$, $z_{\rm opt}$) are projected onto the sky plane, defining the apparent offset $d_{\rm app}$ into the sky plane from the observer's point of view. The solid line in the jet marks $z(v=0.9c)$; the stellar distribution ($r_{\rm c,s}$) is shown in orange. Not to scale.}
    \label{fig:schematic}
\end{figure*}

\begin{figure}
    \centering
    \includegraphics[width=\linewidth]{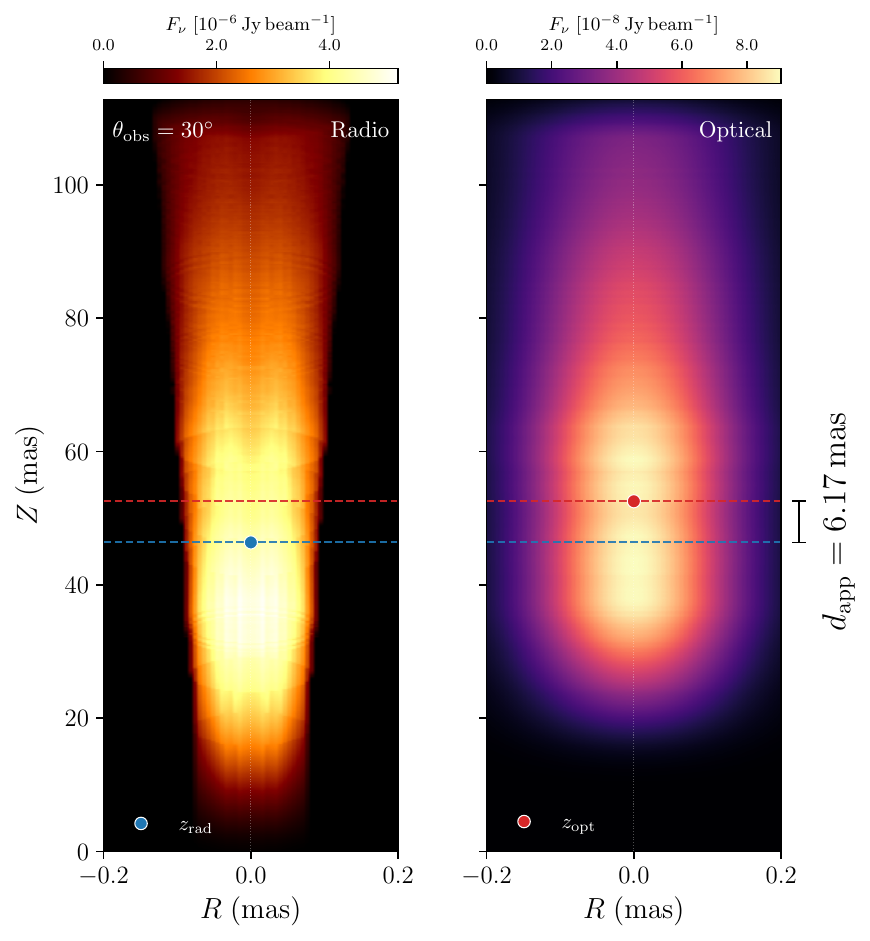}
    \caption{Synthetic synchrotron surface-brightness maps at $43~\mathrm{GHz}$ (\emph{left}) and optical $5\times10^{14}~\mathrm{Hz}$ (\emph{right}) for a representative run with $L_{\rm j}\simeq10^{43}~\mathrm{erg\,s^{-1}}$, $r_{\rm c,s}=500~\mathrm{pc}$, $\theta_{\rm obs}=30^\circ$, at redshift $z=1$. Dashed lines mark the intensity-squared centroid positions $z_{\rm radio}$ (blue) and $z_{\rm opt}$ (red); the bracket to the right of the optical panel indicates the resulting $d_{\rm app}=6.17~\mathrm{mas}$.}
    \label{fig:maps}
\end{figure}

Finally, we generate synthetic radio ($43~\mathrm{GHz}$, chosen as a representative VLBI frequency for parsec-scale AGN imaging) and optical ($5\times10^{14}~\mathrm{Hz}$) synchrotron emission maps by projecting the emergent intensity onto the sky plane. In post-processing, we convolve the maps with representative Gaussian point-spread functions and compute image-plane flux-weighted centroids, from which we define the apparent angular separation $d_{\rm app}$ between radio and optical positions \citep{Plavin_2019}. This procedure provides an idealized, centroid-based estimate of the positional offset; we do not perform visibility-based VLBI forward modeling or simulate group-delay astrometry. To mimic observational conditions, we apply an optical flux threshold consistent with \textit{Gaia} DR3 sensitivity \citep[$F_{\rm AB}\gtrsim10^{-4}\,\rm Jy$ at $3\sigma$;][]{Gaia_2021} and adopt an astrometric precision of $\sim0.1\,\rm mas$. A milliJansky-level radio flux cut has a negligible impact. The resulting $d_{\rm app}$ values are compared with the observed radio-optical offsets following \citetalias{Fichet_2025}. The jet geometry and notation used throughout this work are illustrated in Fig.~\ref{fig:schematic}; an example of the resulting radio and optical surface-brightness maps, with the extracted centroid positions and the corresponding $d_{\rm app}$, is shown in Fig.~\ref{fig:maps}.

\section{Jet dynamics}
\label{sec:dynamics}

\begin{figure*}
    \centering
    \begin{subfigure}{0.49\linewidth}
        \includegraphics[width=\linewidth]{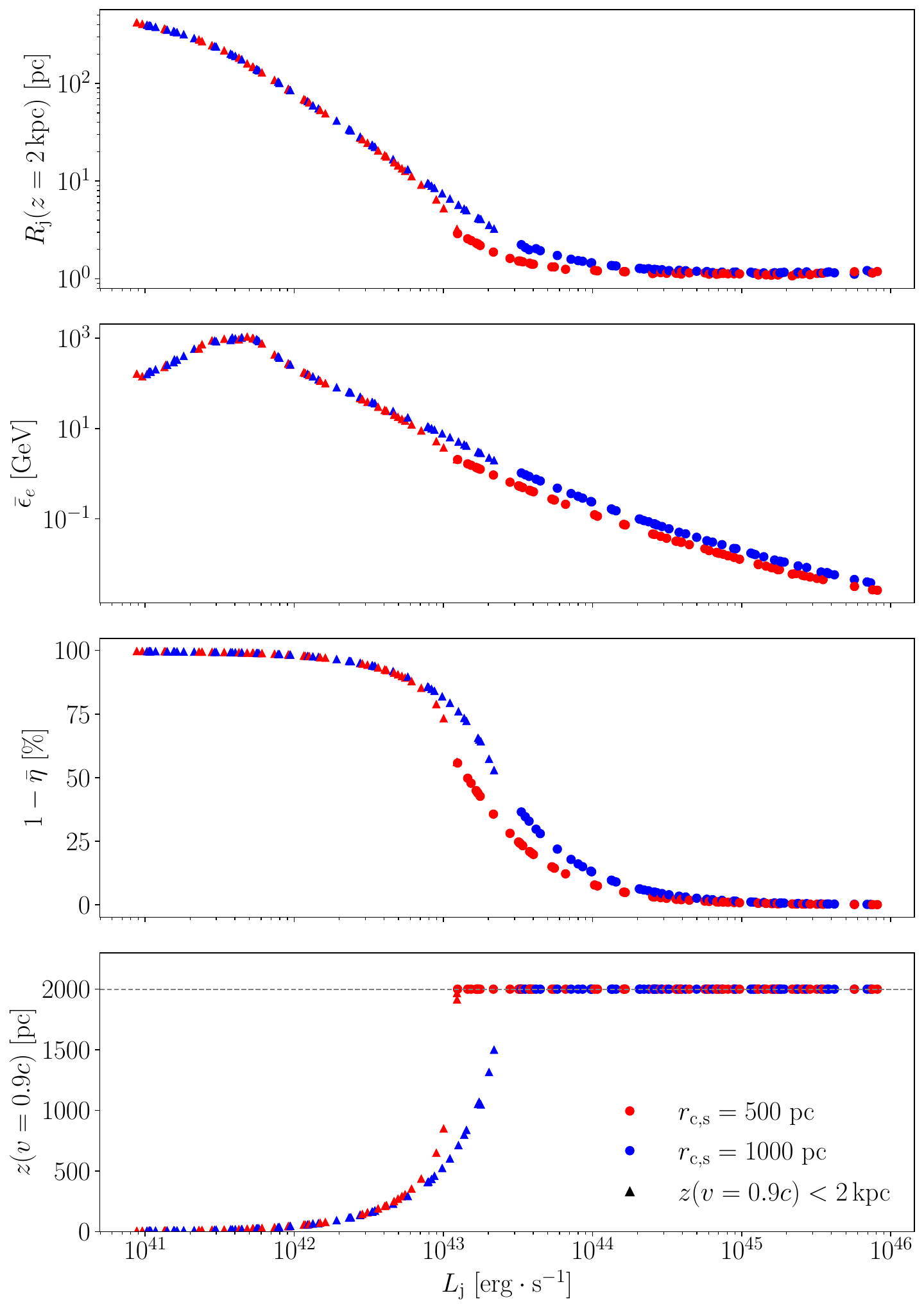}
    \end{subfigure}
    \hfill
    \begin{subfigure}{0.49\linewidth}
        \includegraphics[width=\linewidth]{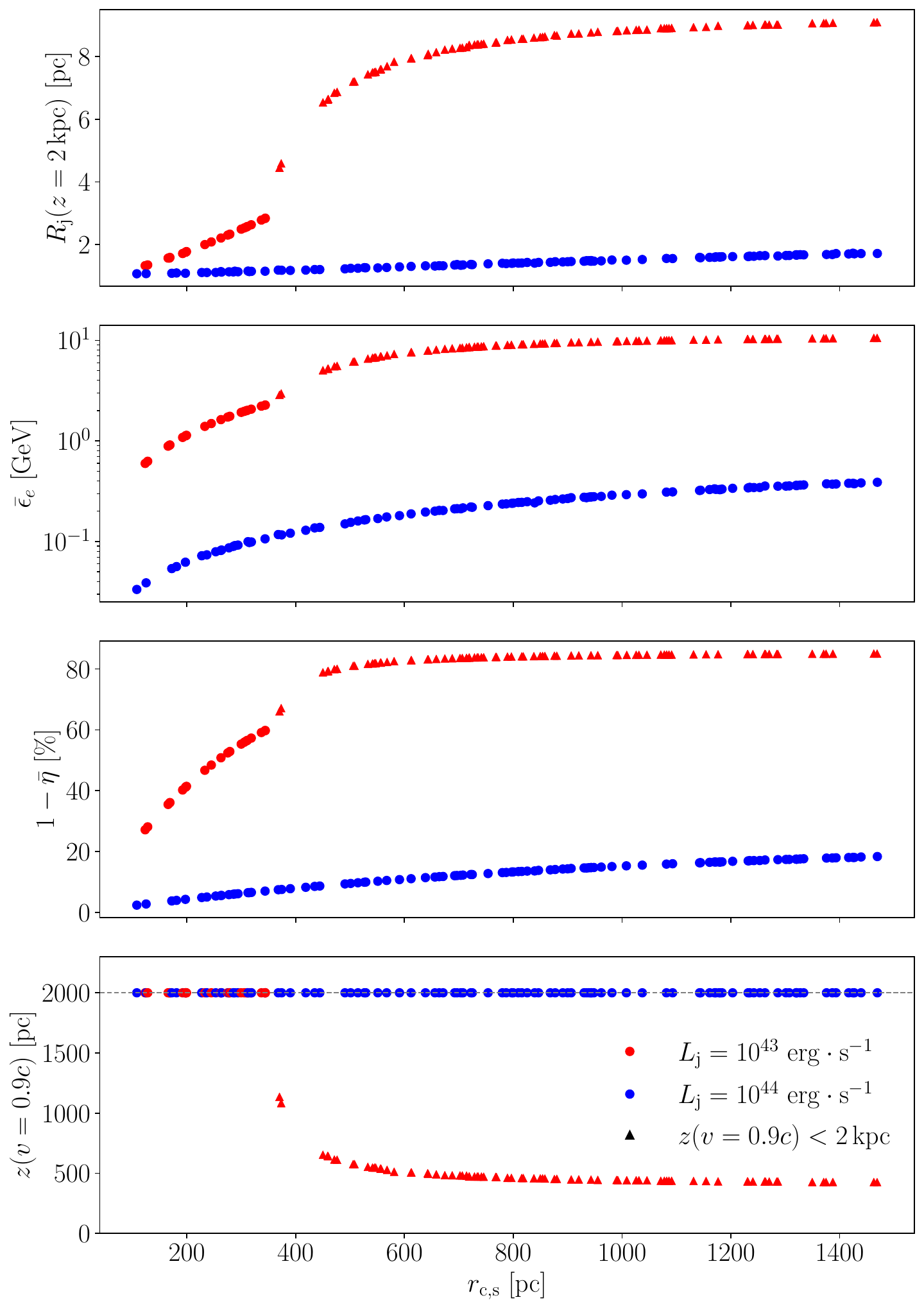}
    \end{subfigure}
    \caption{
    \emph{Left:} Jet radius $R_{\rm j}$ at $z = 2~\mathrm{kpc}$ (\emph{top}), average energy per lepton $\bar{\epsilon}^\prime_{\rm e}$ (\emph{second}), average proton fraction within the jet (\emph{third}), and axial deceleration distance $z(v=0.9c)$ (\emph{bottom}), as a function of the jet power $L_{\rm j}$, for two stellar core radii $r_{\rm c,s}$ (see legend). Each symbol corresponds to one simulation. Circle symbols indicate simulations where the jet remains relativistic throughout the full $2~\mathrm{kpc}$ box; triangle symbols indicate simulations where the jet decelerates before $2~\mathrm{kpc}$, for which the jet radius is obtained by extrapolation (see Sect.~\ref{sec:dynamics}).The dashed line in the bottom panel marks the $2~\mathrm{kpc}$ box limit.
    \emph{Right:} Same as left, but shown as a function of the stellar distribution core radius $r_{\rm c,s}$ for fixed jet powers (see legend).
    }
    \label{fig:dyn}
\end{figure*}

In this section, we investigate how the jet kinetic power, together with the spatial distribution of stellar mass loading, affects jet dynamics. For all simulated jets, we fix the average mass-load rate $Q_{\rm 0} = 5 \times 10^{24}~\rm{g}\,\rm{pc}^{-3}\,\rm{yr}^{-1}$, which is a typical value producing observable radio-optical shifts, as shown in \citetalias{Fichet_2025}.  

To ensure the physical consistency of our analysis, all dynamical quantities are evaluated only up to the axial position $z(v = 0.9c)$, defined as the distance from the jet base at which the bulk flow velocity first drops below $0.9c$. This position sets the upper boundary of the region where the relativistic quasi-one-dimensional approximation adopted in our simulations remains valid. The position $z(v = 0.9c)$ depends strongly on the jet power and on the stellar core radius: for jet powers above $L_{\rm j} \simeq 3 \times 10^{43}~\rm{erg}\,\rm{s}^{-1}$ the jet velocity remains above $0.9c$ throughout the simulation box for any value of $r_{\rm c,s}$. Below $L_{\rm j} \simeq 3 \times 10^{43}~\rm{erg}\,\rm{s}^{-1}$, $z(v = 0.9c)$ moves upstream rapidly for decreasing values of jet power and increasing stellar core radius. For example, at $L_{\rm j} \simeq 10^{43}~\rm{erg}\,\rm{s}^{-1}$, we have $z(v = 0.9c) \sim 500\,\rm{pc}$, for any value of $r_{\rm c,s}$. Below $L_{\rm j} \simeq 10^{42}~\rm{erg}\,\rm{s}^{-1}$, $z(v = 0.9c) < 100\,\rm{pc}$ and the jet opens abruptly near the base. That is why in the rest of the paper, we will distinguish between simulations in which $z(v = 0.9c) \geq 2~\mathrm{kpc}$, corresponding to the full extent of the computational domain, and those in which $z(v = 0.9c) < 2~\mathrm{kpc}$. In the latter case, physical quantities are evaluated within the relativistic region $z \leq z(v = 0.9c)$ and, when required, extrapolated up to $z = 2~\mathrm{kpc}$. These are indicated with triangles in the figures. This approach allows us to compare average jet properties over a common axial distance while preserving the internal consistency of the relativistic flow approximation. When extrapolation is applied, each quantity is treated separately, as detailed below. Specifically, over the fitting interval $[0.2\,z(v=0.9c),\, z(v=0.9c)]$, the jet radius $R_{\rm j}$ is extrapolated by fitting a conical profile $R(z) = s\,z$ forced through the origin by least squares, and evaluating $R\left(2\,\textrm{kpc}\right) =2000\,s~\rm{pc}$. A conical morphology is physically motivated for mildly relativistic, decelerating jets \citep{2014MNRAS.437.3405L, Porth_2015}. The mean electron energy $\bar{\epsilon}^\prime_{\rm e}$ and proton fraction $1-\bar{\eta}$ are each fitted with a power-law $A\,z^{\alpha}$ in log-log space over the same interval, with the slope $\alpha$ determined by the fit, then averaged over the full $0-2~\rm{kpc}$ range. In a conical jet without mass loading, adiabatic losses drive power-law scalings of macroscopic quantities with distance \citep{Kaiser_2006}; in the presence of stellar mass entrainment, the exact index is modified, but we adopt this functional form as a first-order approximation.

To facilitate the interpretation of our results, we study the role of the jet power and the stellar spatial distribution separately. We analyze the impact of these two parameters on three representative quantities: the jet radius at a height of $z = 2~\mathrm{kpc}$ (end of simulation box, top panel in Fig.~\ref{fig:dyn}), the average energy per lepton within the jet (central panel in Fig.~\ref{fig:dyn}), and the average jet baryonic (proton) mass fraction (third panel in Fig.~\ref{fig:dyn}), both averaged within the jet region where $z(v > 0.9c)$, which provide information about the macroscopic structure and microphysical state of the flow. This analysis complements and extends the study run by \citetalias{Fichet_2025} for jets with $L_{\rm j} = 10^{43}~\mathrm{erg}\,\mathrm{s}^{-1}$. 

The top panel of Fig.~\ref{fig:dyn} (left) shows the jet radius at $z = 2~\mathrm{kpc}$ as a function of jet power for two values of $r_{\rm c,s}$. For $L_{\rm j} \leq 10^{43}~\mathrm{erg}\,\mathrm{s}^{-1}$, jets undergo strong lateral expansion, reaching transverse radii in the range $\sim 10-300~\mathrm{pc}$, with only a weak dependence on $r_{\mathrm{c,s}}$. In this regime, the flow is strongly affected by pressure imbalance and stellar mass loading, leading to efficient deceleration and pronounced decollimation \citep{Perucho_2014, Angles_2021}. Around $L_{\rm j} \sim 10^{43}~\mathrm{erg}\,\mathrm{s}^{-1}$, the jet radius at $2\,\rm{kpc}$ decreases rapidly with increasing jet power, marking a transition from a mass-loaded, decollimated regime to a more inertia-dominated flow. For higher jet powers, jets remain significantly collimated and exhibit characteristic radii of $\sim 2-3~\mathrm{pc}$ at kiloparsec scales. 

The second panel of Fig.~\ref{fig:dyn} (left) shows the average energy per lepton within the jet as obtained from our post-processing modeling. For the lowest-power jets, it remains around $\sim 1~\mathrm{GeV}$, consistent with dissipation resulting from heavy mass entrainment. The average lepton energy decreases steadily when the baryonic loading becomes progressively less efficient as a dissipation channel for jet powers larger than $10^{43}~\mathrm{erg}\,\mathrm{s}^{-1}$. For instance, the value is $\sim 10^{-2}~\mathrm{GeV}$ at $L_{\rm j} = 10^{45}~\mathrm{erg}\,\mathrm{s}^{-1}$ and approaches the electron rest-mass energy ($\bar{\epsilon}^\prime_{\rm e} \sim m_{\rm e}c^2$) around $L_{\rm j} = 10^{46}~\mathrm{erg}\,\mathrm{s}^{-1}$. This trend reflects the decreasing ability of stellar mass loading to convert a significant fraction of the bulk jet kinetic energy into internal energy of the leptonic population as jet inertia increases.  

The third panel of Fig.~\ref{fig:dyn} (left) shows the corresponding entrained average proton fraction, $1 - \bar{\eta}$, within the jet ($\eta$ being the electron fraction, set to unity at the jet base).
At low jet powers, the jet becomes heavily mass-loaded, with baryonic fractions approaching unity. As the jet power increases, the proton content drops sharply, reaching values $\lesssim 10\%$ for $L_{\rm j} \geq 10^{44}~\mathrm{erg}\,\mathrm{s}^{-1}$. This reduction reflects both the reduced relative importance of entrainment and the enhanced inertia provided by the original leptonic jet content. The bottom panel shows the position $z\left(v=0.9c\right)$ where the jet starts to decelerate significantly inside the simulation box. An unperturbed jet will show $z\left(v = 0.9c\right) = 2\,\rm{kpc}$, which is the axial size of our simulation box.

The right panels of Fig.~\ref{fig:dyn} show the same quantities as the left ones as a function of the stellar distribution core radius, for two intermediate jet powers. As in the left panels, the circle symbols indicate whether the jet radius or the other quantities are directly measured at $2~\rm{kpc}$ or averaged over the full jet, while the triangle symbols indicate extrapolation of the radius, and averaged quantities within the jet region where $z(v = 0.9c)$. The abrupt change in slope seen in the top panel around $r_{\rm c,s} \simeq 400~\rm{pc}$ marks the transition from simulations where the jet remains relativistic throughout the full $2~\rm{kpc}$ box (circle symbols, direct measurement) to those that are significantly decelerated, and reach $v = 0.9c$ before $z=2~\rm{kpc}$ (triangle symbols, extrapolated radius). This transition is primarily physical: a more extended stellar distribution, i.e.\ larger $r_{\rm c,s}$, sustains mass loading over a larger axial range, increasing the cumulative momentum transfer and triggering earlier deceleration. This steep drop is visible in the bottom panel of Fig.~\ref{fig:dyn}, which shows $z(v=0.9c)$ dropping sharply from $2~\rm{kpc}$ to $\sim450~\rm{pc}$ within a narrow interval of $\sim80~\rm{pc}$ in $r_{\rm c,s}$. The jet radius data points shown as triangles should therefore be interpreted with care, as they represent an extrapolation from the relativistic region assuming a conical geometry.

The top right panel of Fig.~\ref{fig:dyn} shows that the influence of $r_{\rm c,s}$ on the jet radius is not as strong as that of jet power. We see that jets with powers above $10^{44}~\mathrm{erg}\,\mathrm{s}^{-1}$ are only weakly affected by entrainment from stellar winds, independently of the stellar distribution. At $L_{\rm j} = 10^{43}~\mathrm{erg}\,\mathrm{s}^{-1}$, an increase of $r_{\mathrm{c,s}}$ from $100$ to $1500~\mathrm{pc}$ results in an increase of the jet radius from $\sim 2.5$ to $\sim 8~\mathrm{pc}$. In contrast, significant changes can be observed in both average lepton energy (second panel) and proton fraction (third panel). The former rises from $\sim 10^{-1}$ to $\sim 1~\mathrm{GeV}$, whereas the latter increases from $\sim 25\%$ to nearly unity for a $L_{\rm j} = 10^{43}~\mathrm{erg}\,\mathrm{s}^{-1}$ jet. Both magnitudes reflect sustained dissipation (with the consequent efficient particle acceleration) over larger interaction scales. The plots also indicate that the effects are much weaker for a jet power of $10^{44}~\mathrm{erg}\,\mathrm{s}^{-1}$. 

In summary, the obtained jet dynamics highlight the dominant role of jet kinetic power in setting the jet collimation and composition at kiloparsec scales, with a transition occurring over a narrow jet power range centered around $L_{\rm j} \sim 10^{43.5}~\mathrm{erg}\,\mathrm{s}^{-1}$ for the average mass-loading rate adopted here ($Q_0 = 5 \times 10^{24}~\mathrm{g}\,\mathrm{yr}^{-1}\,\mathrm{pc}^{-3}$). This transition shows only a weak dependence on the stellar spatial distribution. Nevertheless, different stellar populations and mass-loss histories are expected to shift the transition power, as discussed in \citetalias{Fichet_2025}.

\section{Synthetic radiation and radio-optical offsets}
\label{sec:radiation}

\begin{figure*}
    \centering
    \begin{subfigure}{0.49\linewidth}
        \includegraphics[width=\linewidth]{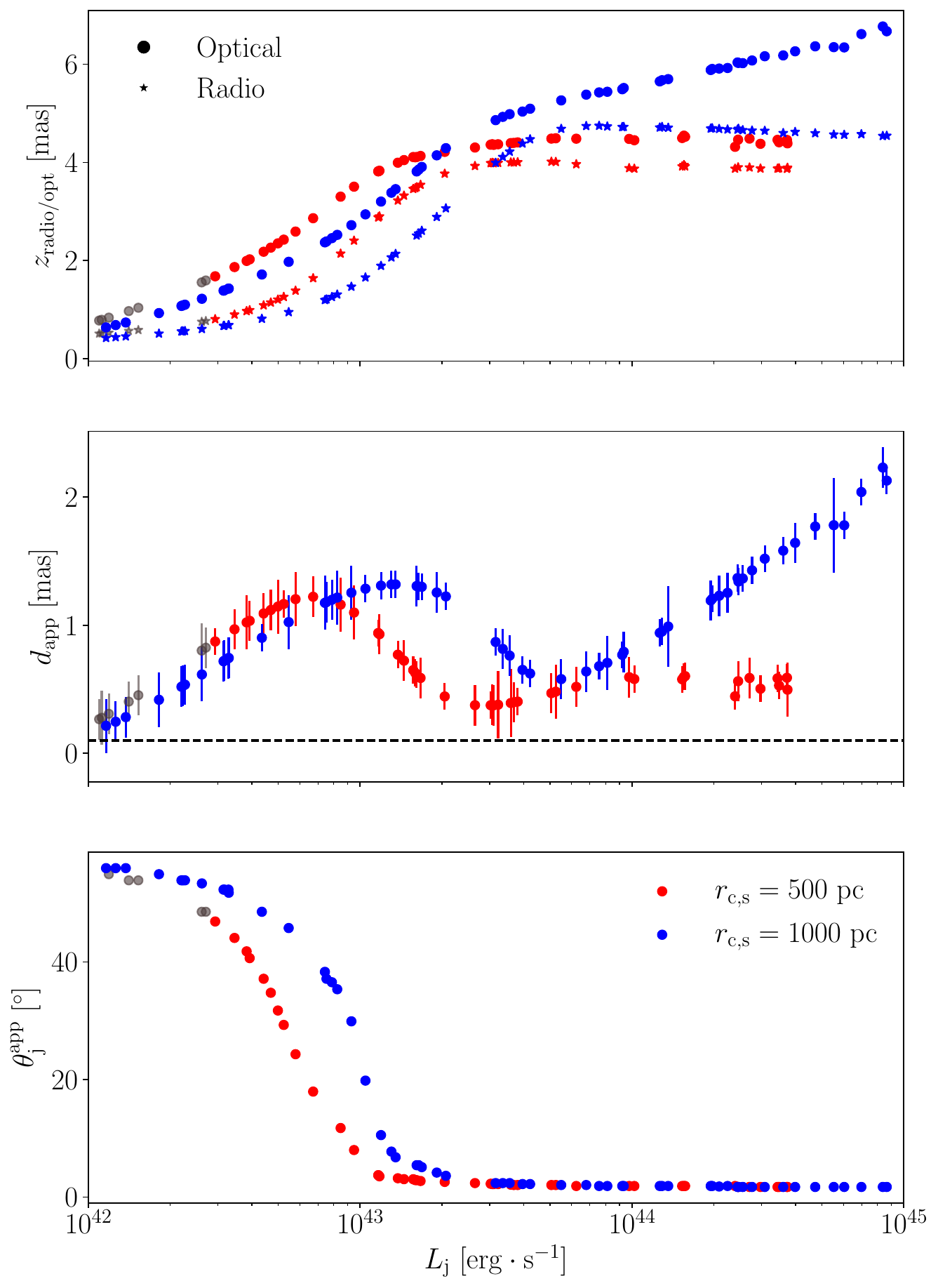}
    \end{subfigure}
    \hfill
    \begin{subfigure}{0.49\linewidth}
        \includegraphics[width=\linewidth]{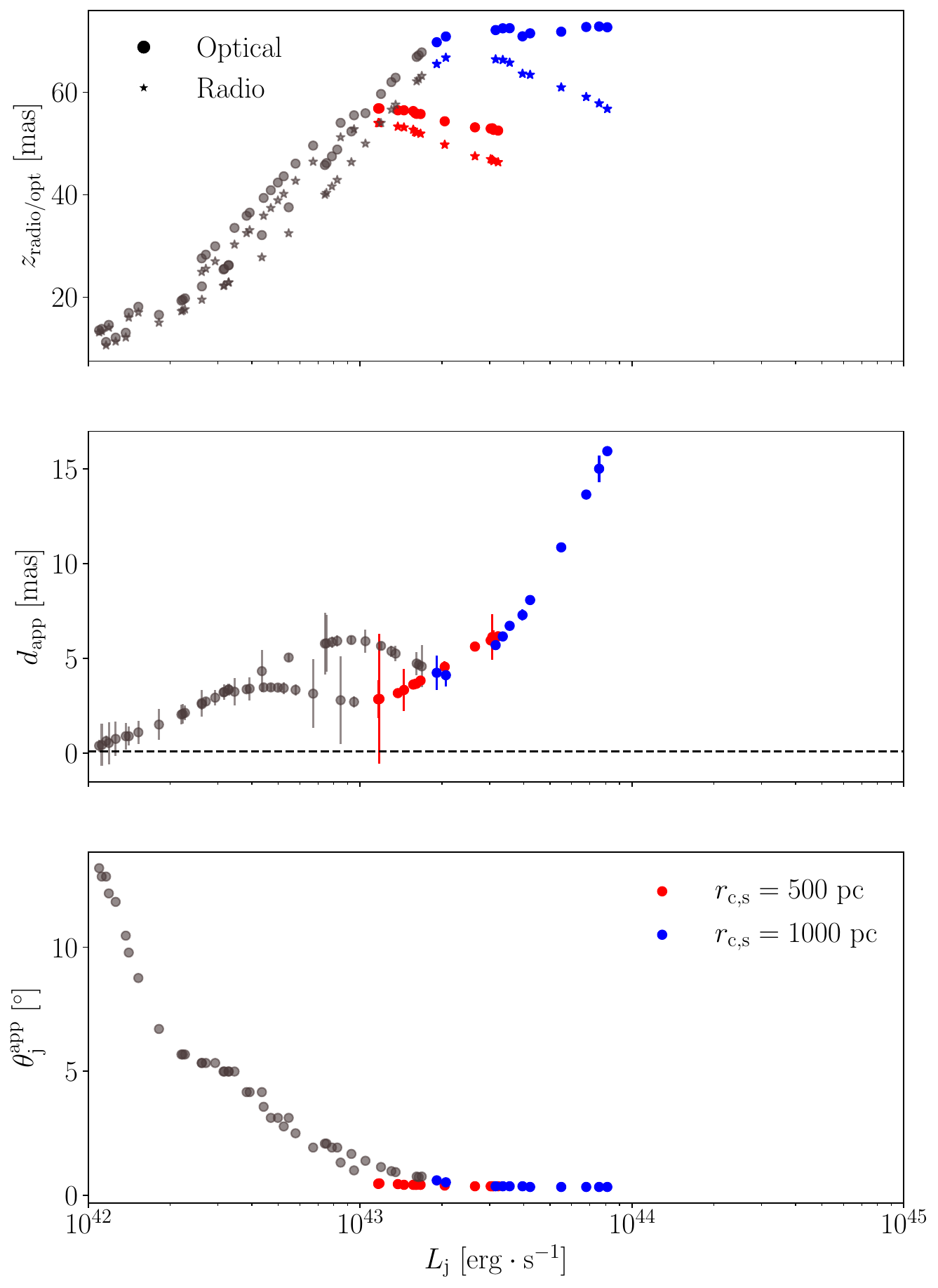}
    \end{subfigure}
    \caption{\emph{Left:} Radio and optical centroid positions ($z_{\rm radio}$ and $z_{\rm opt}$; top), resulting radio-optical offset $d_{\rm app}$ (center, where error bars correspond to the $1\sigma$ uncertainty on the centroid positions, estimated from the effective PSF widths), and apparent jet opening angle $\theta^{\rm app}_{\rm j}$ derived from the $43~\mathrm{GHz}$ emission (bottom), shown as a function of jet power $L_{\rm j}$ for two stellar core radii $r_{\rm c,s}$ (see legend). The dashed line in the central panel represents the \emph{Gaia} astrometrical precision. These results are obtained for a viewing angle $\theta_{\rm obs} = 5^\circ$. Gray symbols indicate cases for which the corresponding centroid lies downstream of $z(v=0.9c)$ and are therefore excluded from the analysis. \emph{Right:} Same as left, but for a viewing angle $\theta_{\rm obs} = 30^\circ$.}
    \label{fig:offset_L}
\end{figure*}

From the synthetic radio ($43\,\rm{GHz}$) and optical ($5\times10^{14}~\rm{Hz}$) surface-brightness maps obtained as described in Sect.~\ref{subsect: riptide}, we extract flux-weighted brightness positions, "centroid" positions, and compute their projected angular separation, hereafter the radio-optical offset $d_{\rm app}$, following \citetalias{Fichet_2025}. For illustration, we adopt $z=1$, corresponding to the median redshift of the \citet{Plavin_2019} sample, and retain only cases where the optical emission exceeds the \emph{Gaia} detection threshold.

These maps also allow us to quantify the apparent jet opening angle $\theta^{\rm app}_{\rm j}$ and to investigate how the spatial morphology of the emission varies with jet power, viewing angle, and stellar distribution. Consistent with the dynamical limitations discussed in the previous section, synthetic emission maps are computed over the full simulation domain, but radio and optical centroid positions (and derived quantities) are only considered physically meaningful when the corresponding centroid lies upstream of $z(v = 0.9c)$. The axial position $z(v = 0.9c)$ is converted into an apparent projected distance from the jet base using the viewing angle $\theta$ and the angular-diameter distance $D_A = D_L/(1+z)^2$, which relates physical sizes to observed angles at cosmological distances. This conversion is purely geometrical: relativistic effects such as Doppler boosting modify the observed emissivity along the jet, but do not shift the apparent position of the kinematic transition in the steady-state approximation adopted here. Results related to radio and optical centroids situated downstream $z(v = 0.9c)$ are reported anyway and are only used for qualitative interpretations.

\subsection{Viewing angle}
\label{subsec: Viewing angle}

The top and central panels of Figure~\ref{fig:offset_L} show the positions of the radio and optical centroids (top) and the resulting offset $d_{\rm app}$ (central) as a function of jet power, for a viewing angle of $5^\circ$ (left), and two stellar core radii, $r_{\mathrm{c, s}} = 0.5$ and $1~\rm{kpc}$.  At low powers, the jets undergo strong deceleration, and the radio and optical centroids are nearly coincident (small offsets; central panel). With increasing jet power, the separation between the two peaks grows, with the optical peak appearing ahead of the radio one (positive offsets). This behavior is common to all stellar distributions. At larger jet powers, the radio-optical offset reaches a similar maximum value for both stellar distributions ($d_{\rm app} \sim 1~\rm{mas}$) but is shifted in jet power ($\sim 7 \times 10^{42}$ and $\sim 2 \times 10^{43}~\rm{erg}\,\rm{s}^{-1}$, for $r_{\rm c,s} = 0.5$ and $1\,\rm{kpc}$, respectively). This maximum value remains above the \textit{Gaia} sensitivity, showing a window where the radio-optical core shift can be more easily detectable for jet powers around $10^{43}~\rm{erg}\,\rm{s}^{-1}$, at redshift $z = 1$. 

Beyond the maxima, the radio–optical offset first decreases and then remains close to the minimum value ($\sim 0.5$ mas) for $r_{\mathrm{c,s}} = 0.5\,\mathrm{kpc}$, whereas for $r_{\mathrm{c,s}} = 1\,\mathrm{kpc}$ it increases for $L_{\rm j} \gtrsim 6 \times 10^{43}\,\mathrm{erg\,s^{-1}}$. This growth in $d_{\rm app}$ reflects the increasing extent of the dissipation region in powerful jets, which shifts the optical peak downstream, while the radio peak remains unchanged because it is set by opacity. In principle, this behavior would make such shifts detectable even in powerful sources. However, the energy dissipation in these jets is too low, causing the optical emission to fall below the sensitivity of \textit{Gaia} ($L_{\rm j} \gtrsim 4 \times 10^{44}\,\mathrm{erg\,s^{-1}}$, for $r_{\mathrm{c,s}} = 0.5\,\mathrm{kpc}$; $L_{\rm j} \gtrsim 8 \times 10^{44}\,\mathrm{erg\,s^{-1}}$, for $r_{\mathrm{c,s}} = 1\,\mathrm{kpc}$).

Figure~\ref{fig:offset_L} (right) shows the same quantities for a larger viewing angle of $30^\circ$. The top panel shows the location of the radio and optical centroids. Doppler (de-)boosting sets the maximum jet power at which the optical centroid remains visible: for instance, it is lost around $3 \times 10^{43}~\rm{erg}\,\rm{s}^{-1}$ for $r_{\rm c, s} = 0.5~\rm{kpc}$ and at $\sim 10^{44}~\rm{erg}\,\rm{s}^{-1}$ for $r_{\rm c, s} = 1~\rm{kpc}$. In this Figure, gray symbols indicate the cases in which the radio and/or optical centroid positions are downstream of the position where the velocity becomes smaller than $0.9c$, which happens for low-power jets. Despite this limitation, we expect similar trends as the ones observed for a smaller viewing angle (see Fig.~\ref{fig:offset_L}, left): a plateau of the radio-optical offset due to strong dissipation closer to the jet base, before it falls to zero for the weakest jets. 

\begin{figure*}
    \centering
    \begin{subfigure}{0.49\linewidth}
        \includegraphics[width=\linewidth]{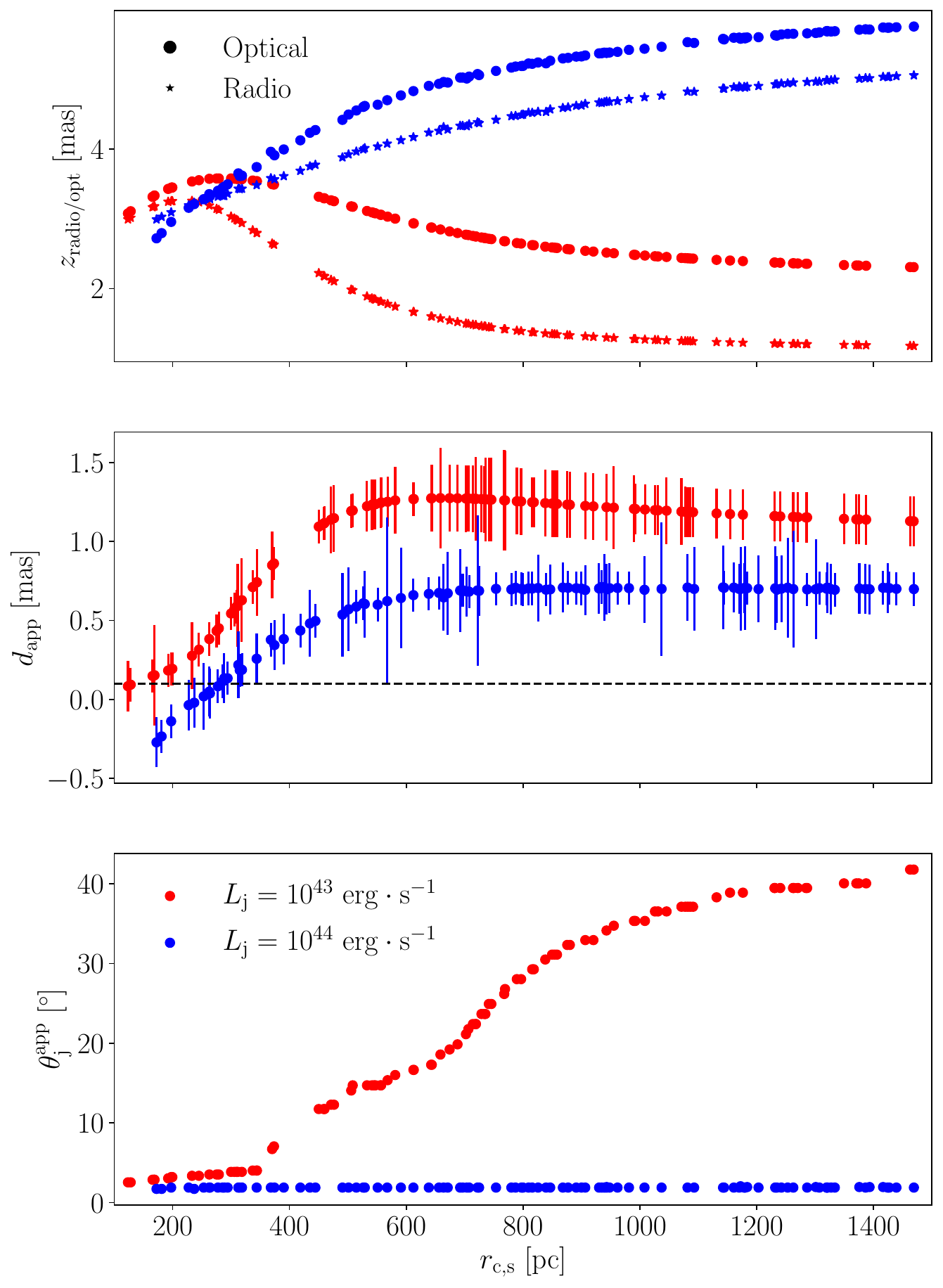}
    \end{subfigure}
    \hfill
    \begin{subfigure}{0.49\linewidth}
        \includegraphics[width=\linewidth]{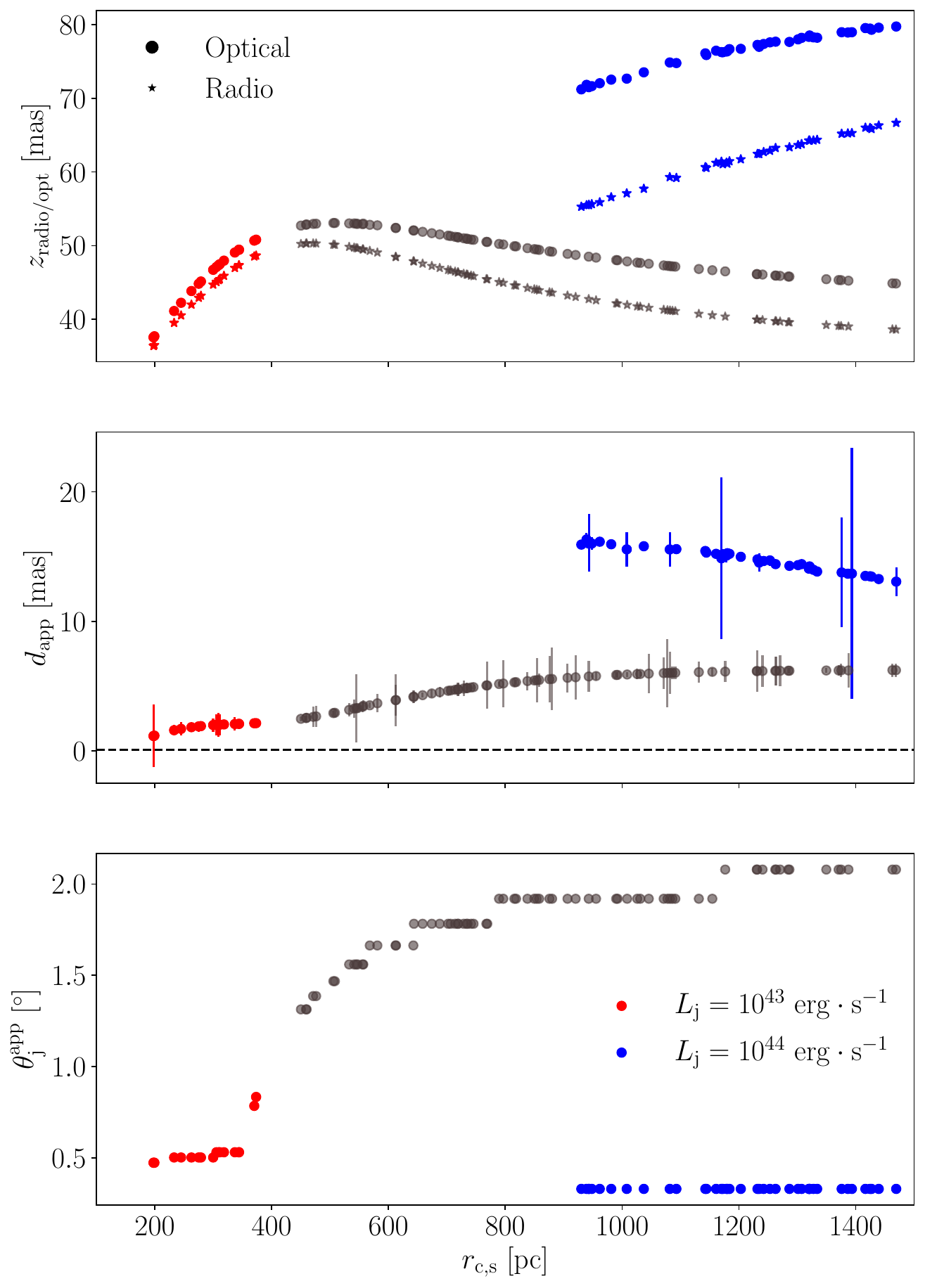}
    \end{subfigure}
    \caption{\emph{Left:} Radio/optical centroid position ($z_{\rm radio}/z_{\rm opt}$, top), the resulting radio-optical offset $d_{\rm app}$ (center, where error bars correspond to the $1\sigma$ uncertainty on the centroid positions, estimated from the effective PSF widths), for two jet powers $L_{\rm j}$ (see legend), and the apparent opening angle from the radio emission (bottom) as a function of the stellar radius $r_{\rm c,s}$ and obtained with $\theta_{\rm obs} = 5^\circ$. The dashed line in the central panel represents the \emph{Gaia} astrometrical precision. If any, gray points represent excluded radio/optical centroids, as their positions are downstream of the region where $v > 0.9c$. \emph{Right:} Same as left, but with $\theta_{\rm obs} = 30^\circ$.}
    \label{fig:offset_rcs}
\end{figure*}

The radio emission maps are used to estimate the apparent jet opening angle $\theta^{\rm app}_{\rm j}$ directly from the projected surface-brightness distribution. See \citet{2017MNRAS.468.4992P} and \citet{2020MNRAS.495.3576K} for details on opening angle measurements from VLBI images. Assuming a conical morphology, we measure the transverse jet width at multiple axial positions within the relativistic region (where $v > 0.9c$), using a fixed fractional intensity threshold of the peak brightness, and derive $\theta^{\rm app}_{\rm j} = 2 \arctan\left[(w/2)/z\right]$ ($w$ being the local jet width, and $z$ the associated axial position) from the maximum projected half-opening angle. This quantity represents a purely image-based, projected opening angle and therefore depends both on intrinsic jet geometry and on the viewing angle. The bottom panels of Fig.~\ref{fig:offset_L} show the resulting $\theta^{\rm app}_{\rm j}$ values derived from $43\,\rm{GHz}$ images at viewing angles of $5^\circ$ (left) and $30^\circ$ (right), respectively, as a function of jet power.

In the first case, we see how mass-load by stellar winds can contribute to decollimation of jets with powers below  $L_{\rm j} \sim 2 \times 10^{43}~\rm{erg}\, \rm{s}^{-1}$, while not significantly affecting more powerful ones. The resulting apparent opening angles reach relatively high values at a viewing angle of $5^\circ$ (up to $50^\circ$) for low-power jets before a sharp decrease around $10^{43}~\rm{erg}\,\rm{s}^{-1}$, while jets remain quasi-cylindrical for larger powers. At viewing angles of $30^\circ$ a similar behavior is observed, although the values given for jets with larger opening angles are based on estimates at $z(v=0.9c)$.

\subsection{Stellar distribution}

To further examine the effect of the stellar core radius on the radio-optical offset and jet collimation, we show the centroid positions (top panel) and corresponding offset $d_{\rm app}$ (middle panel) as a function of $r_{\mathrm{c, s}}$ in Fig.\,\ref{fig:offset_rcs}, for a viewing angle of $5^\circ$ (left), and for a viewing angle of $30^\circ$ (right). The plots display the results for two jet powers: $L_{\rm j} = 10^{43}$ and $10^{44}~\rm{erg} \, \rm{s}^{-1}$. 

At $5^\circ$ (Fig.~\ref{fig:offset_rcs}, left column), only radio-optical offsets for $r_{\rm c, s} \gtrsim 0.2-0.4~\rm{kpc}$ (depending on the jet power) are resolvable by \textit{Gaia} -i.e., appear over the \emph{Gaia} angular resolution indicated by a dashed line in the plot. In contrast, at $30^\circ$ (Fig.~\ref{fig:offset_rcs}, right), all offsets are resolvable, but not all of them are detectable or lie within the region where $v > 0.9c$: for instance, for a $10^{44}~\rm{erg}\,\rm{s}^{-1}$ jet, offsets can only be detected when $r_{\rm c, s} > 1~\rm{kpc}$. At a jet power of $10^{43}~\rm{erg}\,\rm{s}^{-1}$, and $r_{\rm c,s} > 0.4~\rm{kpc}$, radio and/or optical centroids are situated downstream of our validity region, and are therefore not considered. Nonetheless, we expect a similar trend to the one derived at $\theta_{\rm obs} = 5^\circ$, i.e., an increase of the offset before reaching a local maximum followed by a plateau or a smooth decline. 

The dependence of the apparent jet opening angles with $r_{\mathrm{c, s}}$ is shown in the bottom panels of Fig.\,\ref{fig:offset_rcs}. At low jet powers, the opening angle increases with $r_{\mathrm{c, s}}$, reflecting stronger dissipation, in contrast to the case of more powerful jets, which remain basically cylindrical.

\subsection{Summary of the results}

In summary, there is a clear entanglement between the roles of the jet power, stellar distribution, and jet orientation on the values obtained for the radio-optical offsets. 

For small observation angles, in general, smaller jet powers and enhanced mass-load (via larger stellar distribution cores) are required for the offset to be observable. For larger viewing angles, offsets suffer from detectability rather than resolvability (see also \citetalias{Fichet_2025}). We observe that an extended stellar distribution core could compensate for an, in principle, limited degree of dissipation in the case of powerful jets. The radio-morphological transition from large to low opening angle jets occurs at comparable powers of about $L_{\rm j} \sim 10^{43}~\rm{erg}\,\rm{s}^{-1}$, and is reflected in the jet radii obtained at $2\,\rm{kpc}$. Although we have not included all the possible dissipative mechanisms in our limited quasi-one-dimensional simulations, this result gives us the correct order of magnitude of jet powers at which large-scale morphologies transit from FR\,II to FR\,I. The inclusion of extra dissipative processes would probably enhance the differences between the two regimes found here, but we do not expect that this would change the jet power interval that we find for the transition.

Our results indicate that, at cosmological distances (e.g., $z \sim 1$), the optical jet emission is unlikely to remain sufficiently bright relative to the nucleus to produce measurable radio-optical centroid offsets once the jet reaches the strongly dissipative regime characteristic of FR\,I jets for $\theta_{\rm obs} > 5^\circ$ \citep{2014MNRAS.437.3405L}. This does not contradict the presence of prominent optical jets in nearby sources such as M\,87, where spatially resolved observations reveal extended emission powered by additional large-scale dissipation mechanisms (e.g., standing shocks such as HST-1) that are not included in our present modeling.

\renewcommand{\arraystretch}{1.4}

\section{Discussion}
\label{sec:discussion}

Under the assumption that a fraction of the dissipated energy is invested into particle acceleration, our jet simulations show that detectable milliarcsecond radio-optical separations arise within a restricted range of jet kinetic powers, namely $10^{42} \lesssim L_{\rm j} \lesssim 10^{44}~\rm{erg}\,\rm{s}^{-1}$. At low jet powers or large viewing angles, a significant fraction of the simulated jets lie beyond the axial position $z(v=0.9c)$, where our quasi-one-dimensional treatment ceases to be reliable (see Fig.\,\ref{fig:dyn}), thereby limiting the range of offsets that can be robustly interpreted. In such cases, the trends identified in this work should be regarded as indicative of the underlying physical behavior rather than as precise quantitative predictions.

Combined with the offset distributions modeled in \citetalias{Fichet_2025}, and considering only the impact of stellar mass loading within the jet, we can draw two main conclusions: (i) dissipation induced by stellar mass loading constitutes a viable physical mechanism for particle reacceleration and for producing positive radio-optical offsets; and (ii) variations in jet power, stellar distribution properties, and viewing angle systematically shift the location of the centroids along the jet and modulate both its amplitude and detectability. The evolution of these parameters could imprint redshift-dependent trends on the observed radio-optical offset population, making this phenomenon a sensitive probe of jet-galaxy interactions across cosmic time. 

We stress that the specific jet-power ranges discussed in terms of core offset detectability are not universal but depend on the adopted mass-loading normalization $Q_0$ and on redshift. Varying either parameter primarily results in a systematic shift of the same qualitative behaviors toward higher or lower jet powers, rather than in a change of the underlying trends. In this sense, these results should be interpreted as a representative slice of a broader parameter space. In \citetalias{Fichet_2025}, which explicitly explored the dependence of radio-optical offsets on $Q_0$, redshift, and viewing angle, we provided a framework to extrapolate our findings beyond the specific choices adopted here and to connect them across cosmic time.

Finally, although we have established a common initial jet Lorentz factor for all our models, we do not expect significant differences in our results. \citet{Angles_2021} showed that the deceleration (and dissipation) distance was very similar, regardless of the injection Lorentz factor, within the expected values in this region, i.e., $5-10$ (see, e.g., Fig.~5 in that paper).

\subsection{Implications for the FRI/FR\,II dichotomy}

We have seen in Section~\ref{sec:radiation} that there is a clear transition from basically cylindrical to large opening angle jets around $L_{\rm j} \sim 10^{43}~\rm{erg}\,\rm{s}^{-1}$. This probably reveals the jet powers at which direct mass-load by stellar winds may play a significant role in jet deceleration \citep[see also][]{Perucho_2014,Angles_2021}.

If we assume that canonical FR\,I jets studied in, e.g., \citet{2014MNRAS.437.3405L} have jet powers around $10^{43}\,\rm{erg}\,\rm{s}^{-1}$, we see that the apparent opening angles inferred at a viewing angle of $30^\circ$ (see, e.g., Fig.~\ref{fig:offset_L}, right) are systematically smaller than those measured. However, we derive similar opening angles to those reported in the aforementioned paper for smaller jet powers. 

This difference is expected for several reasons. First, our estimates are restricted to milliarcsecond scales and, by construction, only include regions of the flow that remain dynamically valid within our modeling framework, i.e., upstream of the axial position $z(v=0.9c)$. Furthermore, we see that for a viewing angle of $30^\circ$ (more typical of radio-galaxies), a non-negligible fraction of the optical peaks are located beyond $z(v=0.9c)$. 
As a result, the opening angles reported here characterize the still-collimated, mildly decelerated jet rather than the fully developed flaring region where velocities are reported to be significantly smaller \citep[$\simeq 0.3 - 0.4\,c$][]{2014MNRAS.437.3405L}. 

Second, our synthetic maps probe the jet emission at $43~\mathrm{GHz}$, which fades rapidly with distance due to adiabatic expansion losses, in contrast with the $5-8~\mathrm{GHz}$ VLA observations used by \citet{2014MNRAS.437.3405L}. At arcsecond scales, the combined effects of an ambient pressure profile, jet deceleration, and enhanced dissipation lead to non-adiabatic evolution and substantial lateral expansion, producing the large opening angles characteristic of FR\,I flaring regions. These processes are largely associated with the entrainment of external material at the jet boundaries, driven by mechanisms such as small-scale instabilities or the direct interaction of stars with the jet surface \citep[e.g.,][]{Matsumoto_2017, Gourgoliatos_2018, Perucho_2020}. Such boundary-driven entrainment and large-scale geometric flaring cannot be captured within the quasi-one-dimensional, steady-state framework adopted here, which is instead designed to follow the internal evolution of the jet spine under distributed mass loading. Accurately modeling the transition to the flaring region would therefore require fully two- or three-dimensional simulations, which we defer to future work. 

\subsection{Implications on the jet power and stellar properties}
\label{subsec: implications on the jet power and stellar properties}

Our simulations show that each parameter of the model has a significant impact on the radio–optical centroid positions, as well as on the amplitude ($d_{\rm app}$) and detectability of the resulting offsets. In particular, increasing the jet kinetic power $L_{\rm j}$ shifts both radio and optical centroids downstream, increases $d_{\rm app}$, but reduces detectability due to less efficient dissipation. It should be noted that these trends correspond to a controlled exploration of the parameter space in which individual quantities are varied independently. In reality, the model parameters are intrinsically coupled, and the resulting observables reflect the combined effects of jet power, mass loading, and viewing geometry. As a consequence, these trends may develop more complex, non-monotonic behaviors when multiple parameters vary simultaneously, as shown in Figs.~\ref{fig:offset_L}--\ref{fig:offset_rcs}. In particular, the complex evolution of the radio–optical offsets with the jet power can be understood as a direct manifestation of these coupled dependencies, arising from transitions between regimes dominated by dissipation, jet inertia, and radiative or centroid weighting effects. 

At the redshifts considered, the main stellar contribution to mass-load could be the thermally pulsing asymptotic giant branch (TP-AGB) stars, which are characterized by average mass-loss rates between $\dot M \sim 10^{-9}- 10^{-5} \,M_\odot\,\rm{yr}^{-1}$ and low wind velocities, $\sim 10\ \rm{km}\,\rm{s}^{-1}$ \citep{Leitner_2011, Hofner_Olofsson_2018, Oshea_2025}. Their average mass is expected to be, on average, of the order of several solar masses \citep{Herwig_2005}. If we consider a conservative AGB stellar density of $n_{\rm AGB} \sim 10^{-3}\,\rm{pc}^{-3}$ \citep{Oshea_2025}, we derive $Q_0$ values between $2 \times 10^{21}$ to $2 \times 10^{25}~\rm{g}\,\rm{yr}^{-1}\,\rm{pc}^{-3}$, which includes the value used in this work and in \citetalias{Fichet_2025}. 

Concerning the radial extent of the core of the gas and stellar distributions, $r_{\rm c,s}$, it gives an idea of the characteristic interaction/deceleration scale. This value can be estimated using fits to Nuker profiles \citep[see, e.g.,][]{Lauer_2007}, and it should be correlated with the characteristic deceleration distance in low-power jets. In fact, at these redshifts, early-type galaxies have a typical stellar mass $\sim 10^{11}~M_\odot$ and a median effective radius (the radius enclosing most of the stellar mass) between $2$ and $3~\rm{kpc}$ \citep{vandelwel_2014}, which is consistent with our assumptions and the comparison made between our results and \citet{Plavin_2019}. 

In summary, our proposed model to explain the radio-optical offsets detected in the quasar population relies on the interaction between jets and the bulge stellar component of the host. Although quasars might be fed by powerful jets, if they propagate through a relatively extended and dense stellar distribution, i.e., maximizing both $r_{\rm c,s}$ and $Q_0$, an intense process of dissipation within that region, together with a small viewing angle, can bring the optical centroids close to the base of the radio jet. In this case, the effect of cosmological distance on the detectability is counterbalanced by the Doppler boosting, which matches with the fact that the sample in \citet{Plavin_2019} naturally selects Doppler-boosted objects. 

\subsection{Local Universe ($z < 1$)}
\label{subsec: local universe}

The predominance of quasars in radio-optical offset samples at $z\sim1$ could simply reflect the cosmic evolution of black-hole accretion, with luminous, high-Eddington AGN peaks populating intermediate redshifts \citep{Shankar_2009}, in contrast to low-accretion systems such as BL Lac objects and Seyfert galaxies dominating at late cosmic times \citep{Heckman_2014}. Neither the angular resolution nor the received fluxes are, in principle, limiting factors for offset detection at low redshifts.

That being said, massive ellipticals at $z < 1$ are typically dominated by old stellar populations \citep{Thomas_2005}. In such systems, stellar mass return is primarily driven by evolved low- and intermediate-mass stars in the red giant and TP-AGB phases. Although TP-AGB stars dominate the mass-load budget, their mass-loss rates decline significantly with age \citep{Oshea_2025}. Consequently, old elliptical galaxies would provide lower cumulative mass-loading into their jets. 
Regarding jets, Seyferts are also known to host low-power jets (typically $<10^{44}~\rm{erg}\,\rm{s}^{-1}$) with relatively large viewing angles ($\theta_{\rm obs} \leq 30^\circ$ for Seyfert\,II type), whereas BL Lacs show higher intrinsic kinetic power jets (typically between $10^{43}$ to $10^{45}~\rm{erg}\,\rm{s}^{-1}$) with smaller viewing angles \citep[$\theta_{\rm obs} \leq 5^\circ$, see, e.g.,][]{Urry_1995, Heckman_2014, Ghisellini_2014}. 

We can discuss the significant radio-optical offsets found in these two types of galaxies in the frame of our results. Seyfert galaxies frequently contain circumnuclear molecular gas reservoirs, as revealed, for example, by ALMA observations in NGC\,1052 \citep{Kameno_2020}. This gas can either fuel new intermediate-age star formation or be directly entrained by the jet through cloud shredding, effectively increasing the mass-loading rate $Q_0$. In the context of our simulations, even a modest increase in $Q_0$ can move a source into the detectable offset regime. In Seyfert\,I galaxies, where the accretion disk is directly visible, cases in which mass loading is insufficient to trigger efficient energy dissipation will naturally result in disk-dominated optical emission, producing negative radio-optical separations, consistent with the upstream offsets reported by \citet{Plavin_2019}. This interpretation is further supported by \citet{Plavin_2026}, who showed that optical flares in blazars systematically induce Gaia centroid motions: during flux rises, the optical centroid shifts upstream, while during flux decay, it drifts downstream. They localize the flaring region to within $<1$ mas of the VLBI core, corresponding to a few parsecs in projection. In the context of our framework, this behavior naturally complements the negative radio-optical separations expected in Seyfert\,I systems when the accretion disk dominates, while also suggesting that compact jet flares near the radio core may temporarily reduce or even reverse the downstream offsets produced by distributed dissipation. In contrast, in Seyfert\,II galaxies, the accretion disk is obscured, and any detectable radio-optical offset must arise from downstream jet-related emission, leading exclusively to positive separations.

In the case of BL\,Lacs, the Doppler boosting substantially enhances the optical synchrotron emission produced along the jet, making offsets detectable even when stellar winds are weaker. In contrast, our model predicts that higher-power objects such as FR\,II galaxies would rarely produce large downstream offsets, as expected from the irrelevance of dissipation in their case. For instance, Cygnus\,A \citep[with $L_{\rm j}>10^{45}~\rm{erg}\,\rm{s}^{-1}$][]{Snios_2018} displays no detectable VLBI-Gaia displacement. Such jets remain relativistic and emit only weak extended optical radiation outside the core.

\subsection{Early Universe ($z > 2$)}
\label{subsec: early universe}

Above redshifts $z > 2$, the incidence of radio-optical offsets rapidly decreases \citep{Plavin_2019}. Galaxies may not have yet developed significant TP-AGB populations at these early epochs \citep{Bruzual_2003}. Instead, mass-load could be dominated by winds of Wolf-Rayet (WR) stars, with terminal velocities of $v \sim 10^{3}\,\rm{km}\,\rm{s}^{-1}$ \citep{Mokiem_2007} and mass-loss rates as high as $\dot{M} \sim 10^{-4}~M_\odot \, \rm{yr}^{-1}$ \citep{Barlow_1981}. 
However, WR stars are intrinsically rare (only a few thousand are expected in the whole Milky Way) and are preferentially found in galaxies with intense, localized star formation \citep{Liang_2020}. Recent James Webb Telescope observations have confirmed their presence in galaxies at $z \sim 2.2$ \citep{Curti_2025} and even up to $z \sim 6.1$ \citep{Berg_2025}, but these detections remain associated with compact star-forming regions rather than widespread stellar populations.
Consequently, their volumetric number density is expected to be low, implying that WR-jet interactions could only sporadically occur and over limited spatial scales. Such interactions are governed by large-scale, strong shocks rather than gradual mixing from the distributed stellar population \citep[e.g.,][]{Hubard_2006}. The resulting coupling therefore leads primarily to dramatic, localized perturbations, while the efficiency of sustained volumetric mass loading into the bulk relativistic jet is reduced compared to red giants and/or TP-AGB winds \citep[e.g.,][]{Wykes_2015}. Supernova explosions within the jet represent an even more extreme case: the interaction triggers reverse shocks and hydrodynamical instabilities that drive turbulent mixing of the ejecta with the jet material, temporarily decelerating the flow \citep{Longo_2025}. At the same time, radio AGN could be intrinsically more powerful at these epochs, with typical jet kinetic powers exceeding $10^{45}~\rm{erg}\,\rm{s}^{-1}$ \citep{Rigby_2015}, which would limit the dissipation induced by mass loading. 

Combined with a strong observational bias toward bright, disk-dominated quasars in \emph{Gaia} samples, these physical conditions suppress significant offsets or favor negative values at high redshifts: the optical jet emission typically falls below the \emph{Gaia} sensitivity, causing the measured optical centroid to be dominated either by the inner jet (and suppress offsets) or even the accretion disk (resulting in a negative offset) rather than synchrotron emission from the jet. 

\section{Conclusions}
\label{sec:Conclusions}

We have investigated the physical origin and detectability of parsec-scale radio-optical offsets in AGN jets using axisymmetric RMHD simulations with continuous stellar mass loading, post-processed with the RIPTIDE radiative transfer code. We assume that a fixed fraction of the dissipated energy, $\epsilon_{\rm e} \sim 0.1$, is transferred to the acceleration of non-thermal particles. Our results indicate that detectable VLBI-Gaia separations of typically $\sim 0.1-4~\rm{mas}$ (corresponding to projected distances of a few to a few tens of parsecs at $z=1$) arise only within a restricted region of jet-environment parameter space, governed by the interplay between jet kinetic power, stellar mass-loading efficiency, and viewing geometry. Importantly, radio-optical offsets are flux-weighted astrometric signatures rather than purely geometric tracers of dissipation and therefore depend on the relative optical contributions of the jet, accretion disk, and host galaxy. The offsets reported in this work are derived by restricting the analysis to centroid positions located upstream of the axial position $z(v=0.9c)$, beyond which our quasi-one-dimensional treatment ceases to be reliable. As a consequence, the trends identified here characterize the inner, mildly decelerated jets, and the estimates of centroid positions should be regarded as lower limits when dissipation extends farther downstream, particularly at large viewing angles or for extended stellar distributions. Nevertheless, we find those trends to be physically meaningful and useful to derive clear conclusions from our work.

For the stellar environments explored here, intermediate-power jets ($L_{\rm j}\sim10^{42.5}$-$10^{44}~\rm{erg}\,\rm{s}^{-1}$) embedded in hosts with substantial TP-AGB wind densities ($Q_{0}\sim5\times10^{24}~\rm{g}\,\rm{yr}^{-1}\,\rm{pc}^{-3}$) produce the largest positive (downstream) radio-optical offsets, with apparent separations of $\sim0.1$-$4~\rm{mas}$. Below this power range, dissipation occurs too close to the jet base for the radio and optical centroids to separate significantly, while above it the flow remains largely ballistic and the optical synchrotron emissivity is insufficient to displace the centroid unless aided by unusually dense stellar environments or opacity effects.

The jet-power window identified here as the plausible regime for offset detection is not universal but scales with the adopted stellar mass-loading normalization. As demonstrated in \citetalias{Fichet_2025}, variations in $Q_0$ or redshift shift this window without altering the qualitative trends identified in this work, and the semi-analytical framework of that study provides a natural basis for extrapolating our results across a broader parameter space and cosmic time. Viewing geometry further modulates offset detectability through Doppler boosting and projection effects. This naturally explains why flat-spectrum quasars, BL Lac objects, and partially aligned FR~I or Seyfert jets dominate VLBI-\textit{Gaia} offset samples \citep{Plavin_2019}.

Finally, while stellar mass loading is unlikely to be the sole dissipation mechanism in relativistic jets, it constitutes an unavoidable and persistent source of jet perturbation in galactic nuclei. Adding to the results of previous works \citep[e.g.][]{Perucho_2014, Angles_2021}, we also conclude here that significant jet deceleration produced solely by stellar winds could only be possible for jets with powers $L_{\rm j}\leq 10^{43}~\rm{erg}\,\rm{s}^{-1}$. However, even when subdominant, the effect of stellar winds in jets sets a baseline level of dissipation that can be amplified by additional processes such as the development of instabilities or recollimation shocks. 

Future improvements in both optical and radio astrometry will directly test the parameter regime identified in this work. In particular, astrometric precision approaching the tens of microarcsecond level will be crucial to distinguish small offsets from noise, and better multi-frequency VLBI imaging will help disentangle compact core positions from extended jet structure by reducing structure-dependent systematic errors. The planned next-generation Very Large Array \citep[ngVLA,][]{2018ASPC..517...15S}, with an order-of-magnitude increase in sensitivity and resolution relative to current cm-mm facilities ($1.2-116\,\rm{GHz}$) and long baselines for high-fidelity imaging, will enable more accurate localization of compact jet cores and detailed mapping of jet morphology, thereby refining offset measurements and tests of the predicted dependence on jet power and viewing geometry. 

\begin{acknowledgements}

We thank the anonymous reviewer for the relevant comments and suggestions that helped us improve the manuscript. We also thank Sasha Plavin and Elena Shablovinskaya for valuable comments on the manuscript. This work has been supported by Banco Santander and Universitat de València (International Mobility Grant 2025). This work has been supported by the Spanish Ministry of Science through Grant \texttt{PID2022-136828NB-C43}, and from the Generalitat Valenciana through grant \texttt{CIPROM/2022/49}.

YYK was supported by the MuSES project, which has received funding from the European Union (ERC grant agreement No~\texttt{101142396}). Views and opinions expressed are however those of the author(s) only and do not necessarily reflect those of the European Union or ERCEA. Neither the European Union nor the granting authority can be held responsible for them.
\end{acknowledgements}

\bibliographystyle{aa.bst}
\bibliography{bib.bib}

\begin{appendix}
\section{Numerical resolution convergence}
\label{app:convergence}

To verify that our results are not significantly affected by the spatial resolution of the quasi-one-dimensional RMHD grid, we performed a resolution convergence study on a representative simulation in which a clear radio-optical offset is present. We selected one representative run ($L_{\rm j} \simeq 10^{43}~\rm{erg}\,\rm{s}^{-1}$, $r_{\rm c,s} = 500~\rm{pc}$), which lies well within the power range where $d_{\rm app}$ is detectable and maximal (see Fig.~\ref{fig:offset_L}). The fiducial grid for this run has $800 \times 5000$ cells over $20 \times 2000~\rm{pc}$ ($r \times z$). We reran the simulation at resolution multipliers ranging from $0.1\times$ to $2\times$ the fiducial, corresponding to grids from $80 \times 500$ to $1600 \times 10\,000$ cells, and recomputed the synthetic emission maps and centroid positions with \texttt{RIPTIDE} at each resolution. 

\begin{figure}[h!]
    \centering
    \includegraphics[width=\linewidth]{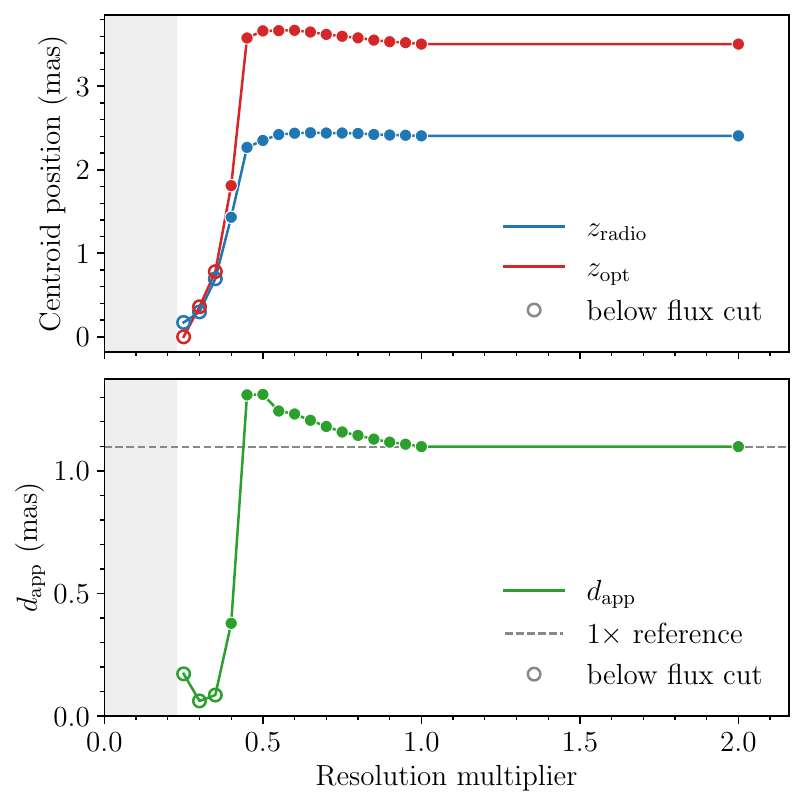}
    \caption{Numerical resolution convergence from one representative run ($L_{\rm j}\simeq10^{43}~\rm{erg\,s^{-1}}$, $r_{\rm c,s}=500~\rm{pc}$, $\theta_{\rm obs}=5^\circ$). \emph{Top:} Radio ($z_{\rm radio}$, blue) and optical ($z_{\rm opt}$, red) centroid positions as a function of the resolution multiplier relative to the fiducial $800\times5000$ grid. Filled circles passed the \textit{Gaia} flux threshold; open circles did not. The light-gray band marks resolutions at which no synchrotron emission is produced. \emph{Bottom:} Corresponding $d_{\rm app}$ values (dark-green circles); the dashed line indicates the fiducial ($1\times$) reference.}
    \label{fig:convergence}
\end{figure}

The results are shown in Fig.~\ref{fig:convergence}. At very low resolutions ($\lesssim 0.25\times$), the grid is too coarse to resolve the jet structure and produces no detectable synchrotron emission (gray shaded region). At intermediate resolutions ($0.3\times$--$0.4\times$), the optical flux falls below the \textit{Gaia} threshold (open symbols), yielding unreliable centroid estimates. From $0.5\times$ onward, both the radio and optical centroids stabilize rapidly (panel~\textit{a}), and $d_{\rm app}$ converges to the fiducial value of $\simeq 1.1~\rm{mas}$ (panel~\textit{b}, dashed line). The $2\times$ run reproduces this value to within $\lesssim 2\%$, confirming that the fiducial resolution is adequate. We therefore conclude that our results are numerically converged and that the radio-optical offsets reported in this work are not resolution artifacts.

\end{appendix}

\end{document}